\documentstyle[preprint,aps]{revtex}
\begin{document}

\title {Confinement and Chiral Symmetry Breaking via
Domain-like Structures in the QCD Vacuum\footnote{ADP-01-25/T459, FAU-TP3-01/6}}

\author{Alex C. Kalloniatis\footnote{akalloni@physics.adelaide.edu.au} }

\address{
Special Research Centre for the Subatomic Structure of Matter ,
University of Adelaide ,
South Australia 5005, Australia}

\author{Sergei N. Nedelko \footnote{nedelko@thsun1.jinr.ru}}

\address{Institut f\"ur Theoretische Physik III,
Universit\"at Erlangen - N\"urnberg ,
Staudtstrasse 7, 
D-91058 Erlangen, Germany
 and Bogoliubov Laboratory of Theoretical Physics, JINR,
141980 Dubna, Russia}

\date{\today}

\maketitle

\begin{abstract} 
A qualitative mechanism for the emergence of 
domain structured background gluon fields
due to singularities in gauge field configurations
is considered, and a model displaying a type of mean field approximation to the 
QCD partition function based on this mechanism is formulated. 
Estimation of the vacuum parameters 
(gluon condensate, topological susceptibility, string constant
and quark condensate) 
indicates that domain-like structures lead to 
an area law for the Wilson loop,
nonzero topological susceptibility and spontaneous breakdown of
chiral symmetry. 
Gluon and ghost propagators in the presence of domains
are calculated explicitly and their analytical properties are discussed.
The Fourier transforms of the propagators  are entire functions and thus
describe confined dynamical fields.
\\
{\bf Pacs: 12.38.Aw 12.38.Lg 14.70.Dj 14.65.Bt 11.15.Tk}
\end{abstract}

\section{Introduction}

Nearly every approach to the problem of the QCD ground  
state accepts that the vacuum is
characterised by strong background gluon fields and,
as results of lattice calculations suggest, 
by a clustered  or lumpy distribution of 
topological charge and action density in configuration space.
We shall refer to such  structures interchangeably 
as ``clusters'' and ``domains''.
They were first observed in typical lattice gauge configurations 
via cooling or smearing algorithms~\cite{Tep85,Ilg86} 
which incrementally suppress quantum fluctuations
by locally minimising or at least reducing the action density.
For more recent work on cooling the reader is referred to \cite{Tep99}.  
On the other hand, the resulting cooled gauge fields tend to give rise to
diminished string constant indicating a loss of confinement.   
An alternative way of analysing the underlying fluctuations of
topological charge density is via the chirality of fermionic modes 
in the background of topological ``lumps'', as originally undertaken 
by \cite{IvN98,EHN98}  and re-discussed recently in~\cite{Hor01}. 
Used with lattice fermions with good chiral properties
such as overlap~\cite{DeGHas01,EH01} or domain wall~\cite{Blum} fermions
the method indicates localisation of
low-lying fermionic modes with definite chirality,
the very modes responsible for the chiral condensate, for
example \cite{BC80}. 
These results can be described in terms of the instanton 
liquid model \cite{Shuryak} and are regarded as evidence for instantons on the 
lattice. While such an interpretation connects clusters of topological charge 
with chiral symmetry breaking, it says nothing about their relevance to 
confinement~\cite{brower}.
However, it might be significant that \cite{EH01} have repeated  
the procedure of \cite{Hor01} with overlap fermions and  no
smearing and still observe strong localisation and definite chirality
of the low lying modes.
This is a piece of evidence for the possibility that localisation of chiral
fermionic modes is due to the effective degrees of freedom responsible for both
confinement and chiral symmetry breaking. 

Several mechanisms of confinement have been proposed
since the formulation of quantum chromodynamics.  All 
try to realise confinement as a dual-Meissner effect,
and thus rely on a condensation of singular gauge configurations 
such as monopoles and vortices \cite{monopoles,ForcDelia}. 
In particular center vortices  are considered 
as effective degrees of freedom relevant both to confinement and chiral
symmetry breaking~\cite{ForcDelia}.  In general, besides the 
above-mentioned configurations
characterised by topologically conserved charges 
there exist also topologically trivial domain wall singularities 
in gauge fixed fields~\cite{jena}.  The form in which singular fields
occur in the gauge fixed formulation varies with the gauge choice, but 
their presence itself is most probably an intrinsically unavoidable feature
of nonabelian theories, universal for a large variety (if not all) gauge 
fixing prescriptions. Consensus about this has been 
growing since the pioneering works of Gribov and Singer~\cite{GrSi}.
This suggests that the manifestation of singular gauge fields 
is linked to the type and dimensionality of the manifold of singularities 
rather than to the peculiarities of their realisation within a particular
gauge fixing prescription. 

We take as a working hypothesis that an effect of this kind 
can be seen in the  restrictive influence of singular gauge fields
on fluctuations in the vicinity of singularities~\cite{lenz1,lenz2}
and formulate a simplified model which allows one to study
manifestations of this effect in vacuum properties and quark-gluon 
dynamics analytically.

The subtleties of separating  
fields into regular and singular parts and the behaviour of regular
fields at the singularities are irrelevant if one could calculate 
the QCD functional integral ``exactly''. 
But these issues   
become crucial if one undertakes approximations~\cite{lenz2}. 
For example, in gauge invariant quantities singularities due to  
ambiguities in gauge fixing should not occur. 
In the action such a finiteness, despite singularities in the gauge
field, occurs either due to cancellations between 
derivative and commutator
parts in the field strength if the singularity is topologically nontrivial
(monopole or vortex) or due to finiteness of both terms separately
for topologically trivial configurations (domain walls).
However this cancellation of singularities in the action density
can be destroyed by unconstrained fluctuations around the singular
fields.  Thus finiteness of the action 
implies specific constraints on fluctuations.  
In Ref.~\cite{lenz2} an example of this is  
considered in detail 
for Polyakov gauge monopoles. 
 
We formulate a model partition function which 
incorporates singularities in gauge fields  
effectively via their restrictive effect on fluctuations.
We assume that singularities are present in general
in gauge potentials, and in their vicinity  
one can divide an arbitrary field, ${\cal A}$, into 
singular $S$ and regular $Q$ parts: 
\begin{eqnarray}
\label{asq-1}
{\cal A}^a_\mu(x)=S^a_\mu(x)+Q^a_\mu(x).
\end{eqnarray}
In order that ${\cal A}$ generates finite action, 
it must be ``close enough'' to a pure gauge configuration in the vicinity of
the singularity, meaning that $[Q,S]=0$ and that  
the field strength for pure gauge $S$ vanishes,
\begin{eqnarray}
\label{smunu}
S_{\mu\nu}=\partial_\mu S_\nu - \partial_\nu S_\mu +ig[S_\mu,S_\nu]=0.
\end{eqnarray}
This can be realised in two ways. 
If Eq.(\ref{smunu})
is satisfied via a cancellation between derivative and
commutator parts then the singularity in $S$ is 
topologically non-trivial and non-abelian.
If the two parts separately vanish then the singularity
is topologically trivial. The gauge potential $S$ is 
then abelian, namely a constant unit colour vector $n^a$ 
can be associated with the field $S$. 

To be explicit, at the cost of generality,  
we shall take the second of these possibilities
and  further assume that singularities in vector 
potentials are concentrated on hypersurfaces $\partial V_j$ ($j=1,\dots,N$) 
in Euclidean space of volume $V$, in the vicinity of which
gauge fields can be divided as above into a sum of a singular pure gauge  
$S^{(j)}_{\mu}$ and regular fluctuation part 
$Q^{(j)}_{\mu}$, with a colour vector $n_j^a$ associated with
$S^{(j)}$.  For such fields  
to have finite action the fluctuations charged with 
respect to $n_j$ must obey specific conditions on $\partial V_j$. 
The interiors of these regions thus constitute ``domains'' $V_j$.
Demanding finiteness of the classical action density, one arrives at 
\begin{eqnarray}
\label{bc}
\breve n_j Q^{(j)}_\mu = 0, \
\psi=-i\!\not\!\eta^j e^{i\alpha_j\gamma_5}\psi,
\
\bar\psi=\bar\psi i\!\not\!\eta^j e^{-i\alpha_j\gamma_5},
\end{eqnarray}
for $\ x\in\partial V_j$,
with the adjoint matrix $\breve n_j=T^an_j^a$ in the condition
for gluons, and a bag-like boundary condition for quarks,
$\eta^j_\mu(x)$ being a unit vector normal to $\partial V_j$.

Eqs.~(\ref{bc}) indicate that 
gauge modes neutral with respect to $n^a_j$ are not restricted
and provide for interactions between domains.  
In a given domain $V_j$ the effect of fluctuations in 
the rest of the system is manifested by an external gauge field 
$B_{j\mu}^a$ neutral with respect to $n_j^a$. This motivates an
approximation in which domains are treated as decoupled but,
simultaneously, with a compensating mean field introduced in their interiors.
The model becomes analytically tractable if we consider
spherical domains with fixed radius $R$ and approximate the 
mean field in $V_j$ by a covariantly constant (anti-)self-dual 
configuration with the field strength
\begin{eqnarray}
\label{meanfield}
&&\hat {\cal B}_{\mu\nu}^{(j)}
=\hat n^{(j)}B^{(j)}_{\mu\nu},
\
\tilde B^{(j)}_{\mu\nu}=\pm B^{(j)}_{\mu\nu},
\
B^{(j)}_{\mu\nu}B^{(j)}_{\rho\nu}=B^2\delta_{\mu\rho},
\nonumber
\\
&&\hat n^{(j)}=t^3\cos\xi_j+t^8\sin\xi_j
\ , \  \xi_j\in\{(2k+1)\pi/6\}_{ k=0}^5,
\end{eqnarray}
where the parameter $B={\rm const}$ is the same for all domains
and the constant matrix $n_j^at^a$ belongs to the Cartan subalgebra,
 the generators $t^a$ being in the fundamental representation.
Note that  since the mean field represents  an effect of fluctuations 
outside the domain
there is no source for this field on the boundary
and therefore it should be treated as strictly homogeneous in all further
calculations. The homogeneity itself appears as a 
simplifying approximation. 
Due to uniformity (on average) of the system outside the domain,
slowly varying fields should be taken into account first of all,   
with leading contributions to this coming from 
covariantly constant fields inside and on the boundary of a domain,   
with a dominance of (anti-)self-dual fields, since they are expected to have 
lower action density~\cite{leutw,mink} than arbitrary constant fields 
(see also the Ginsburg-Landau type consideration in Appendix~B,
where the appearance of a discrete set of values $\xi_j$ is also motivated). 
 
The model for partition function we postulate and then use for calculations   
describes a statistical system 
of finite density $v^{-1}=N/V$ 
composed of $N\to\infty$ noninteracting spherical regions  
in a total Euclidean volume $V\to\infty$, each of which is
characterised by a set of internal parameters with
random values: the angle $\omega_j$ between chromoelectric and 
chromomagnetic fields,
spherical angles $\varphi_j$ and $\theta_j$ of the chromomagnetic field,
the angle $\xi_j$ in the colour matrix $\hat n_j$, 
chirality violating angle $\alpha_j$ entering fermionic boundary condition 
and the coordinate $z_j$ of a domain.
Clusters are characterised also by the fluctuation fields $Q_\mu^{j}$,
$\psi^{j}$ and $\bar\psi^{j}$ satisfying boundary conditions (\ref{bc}),
whose dynamics is driven by the QCD Lagrangian in the presence of the mean 
field.  The propagators of fluctuation fields
for a given background and boundary condition
can be found analytically. Thus this partition function offers a systematic 
prescription for calculation of the correlation functions,
based on a decomposition over fluctuations and 
taking the mean field into account 
explicitly.  Such a treatment of fluctuations as perturbations of a certain 
background field is sensible only if the
essential features of the system can be seen 
in the lowest orders of the decomposition (at least semi-quantitatively).
In other words  one has to verify whether such basic phenomena
as confinement and spontaneous chiral symmetry breaking
are provided by the domain-structured mean field 
and boundary conditions under consideration.  

In the zeroth order of the expansion we shall find
that the gluon condensate, topological 
susceptibility $\chi$ and the string constant $\sigma$ for colour group 
$SU(3)$ take the compact form
\begin{eqnarray}
&&g^2 \langle F^a_{\mu\nu}(x)F^a_{\mu\nu}(x)\rangle=4B^2, 
\
\chi = {{B^4 R^4} \over {128 \pi^2}},
\
\sigma=Bf(\pi BR^2), 
\nonumber\\ 
&&f(z)=\frac{2}{3z}
\left(3-
\frac{\sqrt{3}}{2z}\int\limits_0^{\frac{2z}{\sqrt{3}}}\frac{dx}{x}\sin x
-
\frac{2\sqrt{3}}{z}\int\limits_0^{\frac{z}{\sqrt{3}}}\frac{dx}{x}\sin x
\right),
\nonumber
\end{eqnarray}
while the quark condensate density  at the domain center, 
calculated in the lowest nonvanishing order over quark fields, reads
\begin{eqnarray}
&&\langle \bar\psi \psi\rangle= 
-\frac{q}{2\pi^2R^3(1+q)}
\left[ 2F( BR^2/2\sqrt{3})+F(BR^2/\sqrt{3})\right]
\nonumber\\
&&F(z)=e^z-z-1+\frac{z^2}{4}\int\limits_0^\infty
\frac{dte^{2t-z({\rm coth}t-1)/2}}{{\rm sinh}^2t} 
({\rm coth}t-1),
\
q=B^2R^4/16,
\nonumber
\end{eqnarray}
where $q$  is the absolute value of the topological charge 
associated with a single domain.

To gain numerical estimates of these quantities we fixed the
mean field strength parameter $B$ and domain radius $R$ to fit the 
known value of the string tension,
\begin{eqnarray}
\sqrt{B}=947{\rm MeV}, \ R^{-1}=760 {\rm MeV},
\end{eqnarray}
which leads to the values
\begin{eqnarray}
&&(\alpha_s/\pi)\langle F^2\rangle=0.081 \ {\rm GeV}^4,
\
\sqrt{\sigma}=420 \ {\rm MeV}, 
\nonumber\\
&& \chi=(197 \ {\rm MeV})^4,
\
\langle \bar\psi \psi\rangle= -(228 {\rm MeV})^3,
\end{eqnarray}
with domain charge $q=0.15$ and density  $v^{-1}=42.3 {\rm fm}^{-4}$. 
This estimation shows high density of clusters 
and strong background fields in the system, with confinement of 
static charges and spontaneously broken chiral symmetry. 
There is no separation of scales characterising the 
system, $\sqrt{B}R\approx 1$.
The qualitative picture as well as  numerical values obtained
indicate consistency of the gross features of the model.  

These results suggest that
formation of clusters, predominantly (anti-)self-dual and 
with average size $2R\approx 0.5 {\rm fm}$,  
can have purely quantum origin
whose explanation could require reference 
to the existence of obstructions in gauge fixing.

It should be noted that the physical content of the above numbers can differ
from other approaches.  For instance the QCD sum rules~\cite{SVZ79} 
determination of the gluon condensate is not exactly
comparable to ours, since in our case corrections of order $O(\alpha_s)$ and 
higher contain nonperturbative information via 
explicit dependence of quark and gluon propagators 
on the mean field.

Moreover, the above parameters only give a characterisation of the ``bulk''
properties of the theory and say little about confinement
of dynamical colour modes and hadronisation. Information about
these aspects is contained in the quark and gluon Green's functions,  
in particular in their propagators. 
It is shown that, as expected, the Dirichlet boundary condition
removes gluon zero modes, and the propagator in this problem is well defined,
unlike the analogous problem in the infinite volume.
Both propagators have support in the interior of the
hypersphere, where they have the usual ultraviolet singularity.
Thus at short distances the propagators have standard perturbative form
plus power corrections.
The singularity in the configuration representation of 
propagators is integrable and their Fourier transforms exist, 
so that in momentum space the propagators are entire analytical functions
due to their compact support. This we regard as a manifestation of the 
confinement of dynamical fields.

The paper is organised as follows.  In Section~2 the boundary conditions are 
discussed and the model partition function is defined.
We consider properties of the ensemble of mean fields in Section~3 and 
estimate the lowest dimension gluon condensate,
the topological susceptibility, the string constant  and the quark condensate
in the lowest nonvanishing order in fluctuation fields. 
Gluon and ghost propagators are calculated in Section~4 and their analytical 
properties are discussed.
In Section 5 we give an outline of the problems remaining to
be solved and possible perspectives.
The appendices contain some technical and illustrative material.

\section{The partition function}

In this Section we formulate a partition function
which will be used in subsequent sections for modelling
the QCD  partition function in the presence of clustered background fields.
It should be clear from the very beginning 
that we will not derive the model to be considered from the 
original QCD functional integral.
The mathematically accurate framework for such a derivation, 
a self-consistent mean field approximation requiring 
calculation of the effective action of QCD as 
a functional of the mean field and characteristic functions 
of the domains, is yet to be formulated.   
The best that can be done at this stage is to
identify several ingredients of the formalism 
required for motivating the model within
QCD, postulate the model partition function,  and then look for 
signatures of justification {\it a posteriori}, by means of explicit 
calculations.

Clustered structure of the gauge fields is introduced by 
the proposition that singular configurations
 may not be excluded {\it ad hoc}
from the functional space of integration;
rather
the character of singularities should be restricted by the natural 
requirement that the classical
action density for a given (in general, singular) configuration has to be finite.

We assume that in the vicinity of singularity 
an arbitrary gluon field ${\cal A}$ can be divided as in 
Eqs.~(\ref{asq-1}-\ref{smunu}),
${\cal A}^\mu(x)=S^a_\mu(x)+Q^a_\mu(x)$,
with $Q$ a regular field and $S$ the singular 
pure gauge part,
\begin{eqnarray}
\hat S_\mu(x)=\frac{1}{ig}[\partial_\mu U(x)]U^{-1}(x), \ U(x)=e^{ig\hat f(x)},
\ \hat f=f^at^a.
\nonumber
\end{eqnarray}
The field strength corresponding to $S_\mu^a$ vanishes.
In this paper we will consider abelian singular configurations
\begin{eqnarray}
\partial_\mu\hat S_\nu - \partial_\nu\hat S_\mu=[\hat S_\mu,\hat S_\nu]=0.
\nonumber
\end{eqnarray} 
This can be implemented via  
\begin{eqnarray}
\hat f= \sum\limits_{j=1}^N \hat n_j f_j(x), \ \hat n_j={\rm const}, \ 
[\partial_\mu,\partial_\nu]f_j(x)=0, \ [\hat n_j,\hat n_k]=0,
\nonumber
\end{eqnarray} 
where each of the functions $f_j$ is singular on a 
three-dimensional boundary $\partial V_j$
of the (four-dimensional) region $V_j$, while the matrices $\hat n_j$
belong to the Cartan subalgebra of $SU(3)$ and can be parametrised
by
\begin{eqnarray}
\hat n_j = t_3\cos\xi_j + t_8\sin\xi_j, \ 0\le\xi_j<2\pi.
\nonumber
\end{eqnarray}
The boundaries of the densely packed regions $V_j$ 
necessarily intersect each other and,  for instance,  colour orientation
associated with the boundary becomes ambiguous in the intersection regions. 
Strictly speaking, this means that the abelian singular fields should be 
accompanied by topologically nontrivial vortex-like configurations, such that 
the three-dimensional ``domain wall'' (a topologically trivial object) 
should start and finish at the two-dimensional
singular surfaces, corresponding to a type of dislocation. 
A complete picture would include the whole hierarchy
of singular fields: domain walls, vortices, monopoles and instantons.
It is hard to formulate  a complete approach in a precise way.
A qualitative discussion of this aspect of domain-like structures 
can be found in Ref.~\cite{vBa}.  
Even if we neglect the effects of ``dislocations"  on the boundaries
$\partial V_j$,  a self-consistent consideration is still a complicated problem. 
However, in this case  one can get some idea about features of the
required formalism by means of an artificial example -- QED  
in the presence of the singular background fields,
considered in Appendix A .  
The linearity of electrodynamics enables a
formal definition of the free energy (effective action) as a functional of
a background field and characteristic functions of clusters,
and thus relegates the question about formation of clusters in a typical 
gauge field configuration to a competition between energy and entropy.
In the case of an abelian weakly interacting theory one hardly expects 
domain formation.  
On the contrary in non-abelian strongly interacting theory
singular fields are most probably unavoidable, but unlike QED a straightforward
formulation is  a difficult task.

First of all we should determine the appropriate boundary conditions for the 
fluctuation fields about the singular field $S$ for  finiteness
of the action density. Substituting  Eq.~(\ref{asq-1})
into the QCD Lagrangian, we obtain  
\begin{eqnarray}
\label{lag-qcd}
{\cal L}_{\rm QCD}&=&-\frac{1}{4}Q^a_{\mu\nu}Q^a_{\mu\nu}
+\bar\psi\left[
i\!\not\!\partial - m +g\!\not\!\hat Q +g\!\not\!\hat S
\right]\psi
\nonumber\\
&+&\frac{ig}{2}Q_{\mu\nu}^a
\left[
\breve S^{ab}_\mu Q_\nu^b-\breve S^{ab}_\nu Q_\mu^b
\right]
-\frac{g^2}{2}
\left[ 
(\breve S^2)^{bb'}Q^b_\nu Q^{b'}_\nu -
(\breve S_\mu\breve S_\nu)^{bb'}Q^b_\mu Q^{b'}_\nu 
\right],
\end{eqnarray}
and $Q_{\mu\nu}^a$ is the usual field strength tensor for the fluctuation 
field.
We see from Eq.~(\ref{lag-qcd}) that conditions on the gluon
field arise
\begin{eqnarray}
\label{gbc-1}
\breve n_j Q_\mu = 0 \ {\rm for} \ x\in\partial V_j,
\end{eqnarray} 
while quark fields should satisfy the condition given in Eq.~(\ref{bc}).
Eq.~(\ref{gbc-1}) means that the modes of the gluon field 
longitudinal to the colour vector $n_j^a$ are
not restricted, so it is convenient to decompose gluon fluctuations inside 
the region $V_j$ into transverse and longitudinal parts with respect to $n_j^a$
\begin{eqnarray}
\label{gbc-2}
&&Q^a_\mu = A_\mu^{ja} + n_j^a B_\mu^j, \ n^a_j A_\mu^{aj}\equiv0,
\nonumber\\
&& A_\mu^{ja}=0 \ {\rm for} \ x\in\partial V_j.
\end{eqnarray} 
The separation in Eq.~(\ref{asq-1}) into singular and regular parts 
imposes certain 
restrictions on the gauge transformations if the original and 
transformed fields $Q$  are subject to the same boundary conditions. 
To determine these restrictions it is 
sufficient to consider the infinitesimal transformation
\begin{eqnarray}
&&S_\mu+Q_\mu \to 
S_\mu + Q_\mu + \delta Q_\mu,
\nonumber\\
&&\delta Q_\mu^a=\partial_\mu \omega^a - f^{abc}\omega^b(n^cB_\mu
+A_\mu^c + S_\mu^c),
\end{eqnarray}
from which we conclude that gauge functions should satisfy conditions
\begin{eqnarray}
\label{gfbc}
\omega^a=n^a_j\omega_j + \omega_\perp^{ja}, \ n_j^a\omega_\perp^{ja}=0,
\nonumber\\
\partial_\mu\omega_\perp^{ja}=\omega_\perp^{ja}=0 \ {\rm for} \ 
x\in\partial V_j.
\end{eqnarray} 
The longitudinal functions $\omega_j$ need not be restricted.
The condition Eq.(\ref{gfbc}) dictates that gauge fixing for the fields 
$Q$ should be achieved by means of restricted gauge transformations. 

The original conditions Eqs.~(\ref{bc}) 
show that the interaction of quark and gluon fluctuations    
within the $k-$th region with the field fluctuations
in the rest of the system can be seen as a coupling to external gauge fields 
which are longitudinal to the colour direction $n^a_k$ of the boundary 
$\partial V_k$. 
This feature motivates an approximative treatment of the 
partition function, in which 
clusters are treated as decoupled but, by way of compensation, a  
mean field is introduced in their interior.
A self-consistent mean field approach requires 
calculation of the effective action as 
a functional of the mean field and characteristic functions 
of the domains. Its minima would contain information about 
mean field character, shape and typical domain size.

Here we assume that the effective action favours formation of clusters
with typical size $R$ and nonzero mean field. 
In Appendix~B it is shown that with this 
and an arbitrary constant mean field  the effective action for a domain 
exhibits twelve degenerate
discrete minima corresponding to (anti-)self-dual configurations 
and six values (for $SU(3)$) of the angle $\xi$ associated with the Weyl group. 
There is also a degeneracy in the orientation of the chromomagnetic field.
The value $\xi_0=\pi/6$ is specific for an {\it ansatz} with the 
effective action polynomial in ${\rm Tr}\hat {\cal B}^k$, 
but the period $\pi/3$ is universal.
Since the volume of the domain is finite the degenerate minima 
do not correspond to thermodynamical phases and 
have to be summed in the partition function.   

The partition function for the model is defined as
\begin{eqnarray}
\label{pf-4}
&&{\cal Z}={\cal N}\lim_{V,N}
\prod\limits_{i=1}^N\int\limits_V\frac{d^4z_i}{V}
\int\limits_{\Sigma}d\sigma_i
\int\limits_{{\cal F}^i_Q} {\cal D}Q^i
\int\limits_{{\cal F}^i_\psi}{\cal D}\psi_i 
{\cal D}\bar \psi_i\times
\nonumber\\
&&
\delta[D(\breve{\cal B}^{(i)})Q^{(i)}]
\Delta_{\rm FP}[\breve{\cal B}^{(i)},Q^{(i)}]
e^{
- S_{V_i}^{\rm QCD}
\left[Q^{(i)}+{\cal B}^{(i)},\psi^{(i)},\bar\psi^{(i)}\right]
},
\label{finZ} 
\end{eqnarray}
where the thermodynamic limit assumes $V,N\to\infty$  
with the density $v^{-1}=N/V$ taken finite. 
The fields $Q^{(i)}$, $\psi_i$ and $\bar\psi_i$ are subject to
boundary conditions Eq.(\ref{bc}), in which
the original singularities are effectively encoded.
Interaction between the original domains is
substituted by the mean field.  A background gauge condition is imposed.
The Faddeev-Popov determinant should be calculated on a restricted space of
functions consistent with Eq.~(\ref{gfbc}), which can be implemented in the 
form of integral over ghost fields $(\bar h_j^a, h_j^a)$ subject to 
the boundary condition
 \begin{eqnarray}
\label{ghbc}
\breve n_j h_j = 0 \ {\rm for} \ x\in\partial V_j.
\end{eqnarray} 
The integration measure $d\sigma_i$ is 
\begin{eqnarray}
\label{measure}
&&\int\limits_{\Sigma}d\sigma_i\dots=\frac{1}{48\pi^2}
\int\limits_0^{2\pi}d\alpha_i
\int\limits_0^{2\pi}d\varphi_i\int\limits_0^\pi d\theta_i\sin\theta_i
\times
\\
&&\int\limits_0^\pi d\omega_i\sum\limits_{k=0,1}\delta(\omega_i-\pi k)
\int\limits_0^{2\pi} d\xi_i
\sum\limits_{l=0}^{5}\delta\left(\xi_i-(2l+1)\frac{\pi}{6}\right)
\dots .
\nonumber
\end{eqnarray}
Here $\varphi_i$ and $\theta_i$ are the spherical angles of
the chromomagnetic field,
$\omega_i$ is the angle between the chromomagnetic and
chromoelectric fields, $\xi_i$ is the angle in the colour
matrix $\hat n_i$, $\alpha_i$ is the chiral angle 
and $z_i$ is the centre of the domain $V_i$ with the boundary 
$$
(x-z_j)^2=R^2.
$$

The partition function Eq.~(\ref{finZ}) describes a statistical system 
of density $v^{-1}$ composed of noninteracting hyperspherical 
clusters, each of which is
characterised by a set of internal parameters and
whose internal dynamics are represented by the fluctuation fields.
Correlation functions can be calculated 
taking the mean field into account
explicitly and decomposing over the fluctuations.
First of all we consider vacuum characteristics of the system
to zeroth order in this expansion.

\section{Vacuum Properties to lowest Order in Fluctuations}

The above prescribed perturbative treatment of fluctuations
means in particular that they cannot change vacuum properties
of the system. Thus our immediate task is to test whether the mean field 
itself reproduces the main nonperturbative characteristics of 
the pure gluonic vacuum. To achieve this we have to compute
vacuum expectation values of a number of basic quantities
omitting integration over fluctuation fields.
Thus we will calculate $n$-point connected correlators of
field strength and thereby the corresponding gluon condensates,
string constant and topological susceptibility.

\subsection{Mean field correlators}  

A straightforward application of Eq.~(\ref{pf-4}) 
to the vacuum expectation value of a product of $n$
field strength tensors, each of the form
$$
 B^{a}_{\mu\nu}(x)
=\sum_j^N n^{(j)a}B^{(j)}_{\mu\nu}\theta(1-(x-z_j)^2/R^2),
 $$
gives for the connected $n$-point
correlation function
\begin{eqnarray}
\langle B^{a_1}_{\mu_1\nu_1}(x_1)\dots B^{a_n}_{\mu_n\nu_n}(x_n) \rangle
& =&
\lim_{V,N\to\infty}\sum_j^N\int_V\frac{dz_j}{V}\int d\sigma_j
n^{(j)a_1}\dots n^{(j)a_n}B^{(j)}_{\mu_1\nu_1}\dots B^{(j)}_{\mu_n\nu_n}
\nonumber\\
&\times&\theta(1-(x_1-z_j)^2/R^2)\dots
\theta(1-(x_n-z_j)^2/R^2)
\nonumber\\
&=& B^{n} t^{a_1\dots a_n}_{\mu_1\nu_1,\dots,\mu_n\nu_n}
\Xi_n(x_1,\dots,x_n),
\end{eqnarray}
where the tensor $t$ is given by the integral
$$
t^{a_1\dots a_n}_{\mu_1\nu_1,\dots,\mu_n\nu_n}=
\int d\sigma_j
n^{(j)a_1}\dots n^{(j)a_n}B^{(j)}_{\mu_1\nu_1}\dots B^{(j)}_{\mu_n\nu_n},
$$ 
and can be calculated explicitly using the measure, Eq.~(\ref{measure}). 
This tensor vanishes for odd $n$. In particular, the integral over 
spatial directions is defined by the generating formula
\begin{eqnarray}
\label{gen-f}
\frac{1}{4\pi}
\int\limits_0^{2\pi}d\varphi_j\int_0^\pi d\theta_j\sin\theta_j
e^{iB^{(j)}_{\mu\nu}J_{\mu\nu}}=
\frac{\sin\sqrt{2B^2[J_{\mu\nu}J_{\mu\nu}\pm 
\tilde J_{\mu\nu}J_{\mu\nu}]}}{\sqrt{2B^2[J_{\mu\nu}J_{\mu\nu}\pm 
\tilde J_{\mu\nu}J_{\mu\nu}]}}
\end{eqnarray}
 where the plus and minus correspond to $B^{(j)}_{\mu\nu}$ being
self-dual or anti-self-dual. The translation-invariant function
\begin{equation}
\Xi_n(x_1,\dots,x_n)=\frac{1}{v}\int d^4z
\theta(1-(x_1-z)^2/R^2)\dots
\theta(1-(x_n-z)^2/R^2)
\label{Xi_n}
\end{equation}
can be seen as the volume of the region of overlap of $n$ 
hyperspheres of radius $R$ and centres ($x_1,\dots,x_n$),
normalised to the volume of a single hypersphere
$v=\pi^2R^4/2$,
$$
\Xi_n=1, \ {\rm for} \ x_1=\dots=x_n.
$$ 
It is obvious from this geometrical interpretation 
that $\Xi_n$ is a continuous function and 
vanishes if the distance between any two points
$|x_i-x_j|\ge 2R$; correlations in the background field have finite range
$2R$. The Fourier transform of $\Xi_n$ is then an entire analytical 
function and thus correlations do not have particle interpretation.
It should be stressed that the statistical ensemble 
of background fields is not Gaussian since all connected correlators
are independent of each other and cannot be reduced 
to the two-point correlations. 

As a simplest application of the above correlators we get 
a gluon condensate density which to this approximation is 
\begin{eqnarray}
g^2 \langle F^a_{\mu\nu}(x)F^a_{\mu\nu}(x)\rangle=4B^2.
\end{eqnarray}
Note that the coupling constant is absorbed into the gauge field.

\subsection{Topological Charge and Susceptibility}

Another vacuum parameter which plays
a significant role in the resolution of the $U_A(1)$ problem
is the topological susceptibility 
\cite{Cre77,Wit79,Ven79}.
To define this we consider first the topological charge density
for the colour group $SU(3)$, 
\begin{eqnarray}
Q(x)={{g^2}\over {32 \pi^2}} \tilde F^a_{\mu\nu}(x)F^a_{\mu\nu}(x),
\nonumber
\end{eqnarray}
which in the mean field approximation takes the form
\begin{eqnarray}
Q(x)={{B^2}\over {8 \pi^2}} \sum_{j=1}^N\theta[1-(x-z_j)^2/R^2]\cos\omega_j,
\end{eqnarray} 
where $\omega_j\in\{0,\pi\}$ depending on the duality of the $j$-th domain.
 We thus see that topological charge density is constant in each domain, 
and the sign of this constant is uncorrelated.
For a given field configuration then the topological charge is additive 
$$
Q=\int_V d^4x Q(x)= q(N_+ - N_-), \   q=B^2R^4/16, \  -Nq\le Q\le Nq
$$ 
where $q$ is a `unit' topological charge,
namely the absolute value of the topological charge of a single
domain, and $N_{+}$ $(N_-)$ is the number of 
domains with (anti-)self-dual field, $N=N_+ + N_-$. 
With fixed total number of domains $N$
the probability of finding the topological charge $Q$ in a given 
configuration is given by the distribution
$$
{\cal P}_N(Q)=\frac{{\cal N}_N(Q)}{{\cal N}_N}=
\frac{N!}{2^N\left(N/2-Q/2q\right)!\left(N/2+Q/2q\right)!},
$$
where ${\cal N}_N(Q)$ is the number of configurations with a given charge
and ${\cal N}_N$ is the total number of configurations.
The distribution is symmetric about $Q=0$, where it has a maximum
for $N$ even. 
For $N$ odd the maximum is at $Q=\pm q$. 
We conclude that topological charge averaged over the ensemble
of clusters vanishes. 

The topological susceptibility
\begin{equation}
\chi = \int d^4x \langle Q(x) Q(0) \rangle
\end{equation}
is determined by the two-point correlator of topological charge density, 
which in the lowest approximation reads 
\begin{equation}
\langle Q(x)Q(y)\rangle={{B^4}\over {64 \pi^4}} \Xi_2(x-y),
\end{equation}
and we get 
\begin{equation}
\chi = {{B^4 R^4} \over {128 \pi^2}}. 
\end{equation}

\subsection{Area law for the Wilson loop}

In the same mean field approximation the Wilson loop is given by the integral
\begin{eqnarray}
W(L)=\lim_{V,N\to\infty}\prod\limits_{j=1}^N\int_V\frac{d^4z_j}{V}
\int d\sigma_j\frac{1}{N_c}{\rm Tr}
\exp\left\{i\int_{S_L}d\sigma_{\mu\nu}(x)\hat B_{\mu\nu}(x)\right\},
\nonumber
\end{eqnarray}
where the measure $d\sigma_j$ corresponds to an integral over  
the set of parameters 
\begin{eqnarray}
\left\{z_k,\phi_k,\theta_k,\omega_k,\xi_k\right\}_{k=1}^N
\nonumber
\end{eqnarray}
of the field strength
\begin{eqnarray}
\hat  B_{\mu\nu}(x)=\sum_k \hat n^{(k)} B^{(k)}_{\mu\nu}\theta(1-(x-z_k)^2/R^2).
\nonumber
\end{eqnarray}
Note that path ordering in our case is not necessary since the matrices 
$\hat n^{(k)}$ are assumed to be in the Cartan subalgebra.

Strictly speaking the contour $S_L$ around which the path-ordered
exponential is integrated should be a rectangle whose Euclidean-time 
length should be taken arbitrarily large before the spatial
length. It is for such a contour that one has a strict
interpretation of the behaviour of the exponent in terms
of a static potential \cite{Wil74,Pes83}. However the expectation that
there be an area law is not dependent on the specific geometry
of the contour. In view of the rotational properties of our approximation
to the vacuum fields, it is computationally more convenient to
consider a {\it circular} contour in the $(x_3,x_4)$
plane of radius $L$ with centre at the origin. 
If an area law is
established, as will be the case, the numerical value of the
resulting string constant would not be precisely that corresponding 
to a rectangular contour. However  due to the fact that the loop
must be taken large in order to extract the potential
the difference between a circle
and a rectangle should not lead to radically different values
of the string constant.

To illustrate the steps in the calculation while avoiding
cumbersome formulae we consider here the case of colour group 
$SU(2)$. The details of $SU(3)$ will be given in Appendix C,
though the final result will be quoted below.
For colour $SU(2)$ we have 
$$\hat n^{(k)}=\epsilon^k\tau_3, \ \epsilon^k=\pm 1.$$
The thermodynamic limit $(V,N\to \infty)$ assumes that the subvolume
$$v=V/N=\pi^2R^4/2$$ is fixed. 
Calculation of the trace in colour space leads to the result
\begin{eqnarray}
\frac{1}{2}{\rm Tr}
\exp\left\{i\int_{S_L}d\sigma_{\mu\nu}(x)\hat B_{\mu\nu}(x)\right\}
&=&
\cos\left( \sum_k\epsilon^k B^{(k)}_{\mu\nu}J_{\mu\nu}(z_k)\right),
\nonumber
\end{eqnarray}
where we have denoted
\begin{eqnarray}
J_{\mu\nu}(z_k)=\int_{S_L}d\sigma_{\mu\nu}(x)\theta(1-(x-z_k)^2/R^2).
\end{eqnarray}
Using the properties of the measure of integration over the collective 
coordinates one gets 
\begin{eqnarray}
W(L)=
\lim_{V,N\to\infty}\left[\int_V\frac{d^4z_j}{V}
\int d\sigma_j\frac{1}{2}
\left(e^{i B^{(j)}_{\mu\nu}J_{\mu\nu}(z_j)}
+
e^{-i B^{(j)}_{\mu\nu}J_{\mu\nu}(z_j)}
\right)\right]^N.
\nonumber
\end{eqnarray}
We have exploited here the property that the integral over collective variables
does not depend on the index $j$.
As the contour of the Wilson loop is in the $(x_3,x_4)$-plane, the
only nonzero components of $J_{\mu\nu}$ are 
\begin{eqnarray}
J_{34}=-J_{43}(z)=\int_{S_L}dx_3dx_4\theta(1-(x-z)^2/R^2),
\end{eqnarray} 
and 
\begin{eqnarray}
B_{\mu\nu}J_{\mu\nu}(z)=2B_{43}J_{43}(z)=2E_3J_{43}(z)=2BJ_{43}(z)\cos\theta,
\end{eqnarray} 
where $\theta$ is the angle between chromoelectric field ${\bf E}$
and the third coordinate axis.
Now we can calculate the integral over the spatial orientations of the vacuum 
field
\begin{eqnarray}
&&\int d\sigma_je^{iB^{(j)}_{\mu\nu}J_{\mu\nu}(z_j)}=
\frac{1}{4\pi}\int_0^{2\pi}d\phi\int\limits_0^\pi d\theta_j\sin\theta_j
e^{2iBJ_{43}\cos\theta_j}=\frac{\sin 2BJ_{43}(z_j)}{2BJ_{43}(z_j)},
\nonumber
\end{eqnarray} 
and the Wilson loop takes the form
\begin{eqnarray}
W(L)=\lim_{N,V\to\infty}\left[\frac{1}{V}\int_Vdz
\frac{\sin 2BJ_{43}(z)}{2BJ_{43}(z)}\right]^N.
\nonumber
\end{eqnarray} 
Calculating the integral over $z$ we obtain finally    
\begin{eqnarray}
W(L)&=&\lim_{N\to\infty}\left[1-\frac{1}{N}U(L)
\right]^N=e^{-U(L)}
\nonumber\\
U(L)&=&\frac{\pi^2R^2L^2}{v}
\left(1-\frac{1}{2\pi BR^2}\int_0^{2\pi BR^2}\frac{dx}{x}\sin x
\right)
+\frac{\pi^2}{v}\left(\frac{4}{3}R^3L+\frac{1}{2}R^4\right)
\nonumber\\
&-&\frac{\pi^2(1-\cos2\pi BR^2)}{v(2\pi B)^2}
+\frac{4\pi^2L}{v(2\pi B)^{3/2}}\int_0^{\sqrt{2\pi BR^2}}dx\sin x^2
\nonumber\\
&-&
\frac{\pi^2L^4}{v}
\int_0^{R^2/L^2}ds
\int_{(1-\sqrt{s})^2}^{(1+\sqrt{s})^2}dt
\frac{\sin\left[
BL^2\left(2\varphi-\sin\varphi+s(2\psi-\sin\psi)
\right)
\right]}{BL^2\left(2\varphi-\sin\varphi+s(2\psi-\sin\psi)
\right)},
\nonumber\\
cos\frac{\varphi}{2}&=&\frac{t-s+1}{2\sqrt{t}}, \
\cos\frac{\psi}{2}=\frac{t+s-1}{2\sqrt{st}},
\end{eqnarray}
where the thermodynamic limit ($N,V\to\infty$, $v=V/N=\pi^2R^4/2$)
has been taken.  One can check that $U(L)=0$ when $B\rightarrow 0$, as it should.

In the limit of large Wilson loop $L\gg R$ the behaviour of the first four 
terms in $U(L)$ can be determined by inspection.  
For the last term a slightly more involved calculation gives  
the large $L$ behaviour   
\begin{eqnarray}
- \frac{\pi^2L^4}{v}
\int_0^{R^2/L^2}ds
\int_{(1-\sqrt{s})^2}^{(1+\sqrt{s})^2}dt
\frac{\sin\left[
BL^2\left(2\varphi-\sin\varphi+s(2\psi-\sin\psi)
\right)
\right]}{BL^2\left(2\varphi-\sin\varphi+s(2\psi-\sin\psi)
\right)} \approx 
-\frac{8\pi^2LR^3}{3v}
\nonumber
\end{eqnarray}
with corrections coming at ${\cal O}(R^4)$. 
Thus only the first term in $U(L)$, going like $L^2$,
displays an area dependence. The final result for the string constant  
for $SU(2)$ is:
\begin{eqnarray}
&&W(L)=e^{-\sigma \pi L^2 + O(L)}, \ \ \sigma=Bf(\pi BR^2),
\nonumber\\
&&f(z)=\frac{2}{z}
\left(1-\frac{1}{2z}\int_0^{2z}\frac{dx}{x}\sin x\right).
\nonumber
\end{eqnarray}
For the case of $SU(3)$, as shown in Appendix C,
the function $f(z)$ turns out to be: 
\begin{eqnarray}
f(z)=\frac{2}{3z}
\left(3-
\frac{\sqrt{3}}{2z}\int_0^{2z/\sqrt{3}}\frac{dx}{x}\sin x
-
\frac{2\sqrt{3}}{z}\int_0^{z/\sqrt{3}}\frac{dx}{x}\sin x
\right).
\label{su(3)f}
\end{eqnarray}
It is positive for $z>0$ and has a maximum for $z= 1.55\pi$.
We choose this maximum to estimate the model parameters 
by fitting the string constant to the lattice result,
\begin{eqnarray}
\label{par-val}
\sqrt{B}=947{\rm MeV}, \ R^{-1}=760 {\rm MeV},
\end{eqnarray}
with unit charge $q=0.15$, density  $v^{-1}=42.3 {\rm fm}^{-4}$ 
and the ``observable'' 
gluonic parameters of the vacuum
\begin{eqnarray}
\label{res1}
&&\sqrt{\sigma}=420 \ {\rm MeV}, \ \chi=(197 \ {\rm MeV})^4,
\nonumber\\ 
&&(\alpha_s/\pi)\langle F^2\rangle=0.081 \ {\rm GeV}^4.
\end{eqnarray}
The high density ensures area law dominance  
already at distances $2L\approx 1.5-2\ {\rm fm}$. 

The result for the gluon condensate is larger
than the most recent estimate within QCD sum rules \cite{Nar97}. 
As already mentioned,
our value is not directly comparable to sum rules results
due to differences in the content of $O(\alpha_s)$ corrections. 
What appears to be important is that all these quantities are nonzero,
and their values can be fit to the expected numbers simultaneously.  
 
Obviously, if $B$ goes to zero then the string constant vanishes.
This underscores the
role of the gluon condensate 
in the confinement of static charges. 
On the other hand we can also 
see that if the number of domains is fixed and the thermodynamic limit is 
defined as $V,R\to\infty, N={\rm const}<\infty$, namely if  
the clusters are macroscopically large,
then $W(L)=1$, which indicates absence of a linear potential
between static (infinitely heavy) charges in a purely homogeneous field. 
However this does not mean that heavy quarks ($m_Q^2\gg B$) are not confined
if domains are macroscopically large.
As is shown in Ref.~\cite{efnedkal}, the nonrelativistic potential
is quadratic in the distance between heavy quarks with the coefficient
proportional to $m_Q^{-1}$.

Since we have integrated over background fields exactly the 
role of finite range of correlation functions is hidden in the above 
calculation. In order to see this role explicitly one would need to decompose
the integrand into an infinite series and integrate term by term.
At this step all correlation functions of the 
background field up to infinite order would be manifest. The arguments of 
Refs.~\cite{DoSim} about the crucial importance of a fast decay of correlators
for confinement of static charges  would be seen to apply here via
this representation.

A comment on the values of the parameters $R$ and $B$ appearing 
in our estimation is in order.
We observe that there is no separation of  
the two scales characterising the vacuum. The average strength
of vacuum fields $B$ and the average 
domain size $R$,
are comparable to each other $\sqrt{B}R\approx 1$. 
Neither large domains nor stochasticity   
of background fields are seen here
which {\it a posteriori} justifies the mean
field averaging prescription in the partition function.  
This prescription corresponds to a system 
less ordered than, for instance, a spin glass. Nor does the partition function
represent a heterophase mixture~\cite{slava}, since the condition for 
quasi-equilibrium is not satisfied: one may not think of these clusters 
as droplets of different
thermodynamic phases as they are too small and too transient  
compared to the basic scale of interactions determined in this picture
by the gluon condensate value.  
The mean field in the clusters is singled out not due to a hierarchy 
of scales, but due to certain specific properties: the 
(anti-)self-duality, and the abelian character if dislocations at the 
boundaries are neglected. Homogeneity of the background field appears as an 
approximation.

\subsection{Quark Condensate Density at Domain Centre}
A complete consideration of the fermionic eigen-problem for the background 
field and boundary conditions under consideration will be given in a separate 
work.  However to complete the picture of vacuum properties in the model,
we estimate here the quark condensate density at the domain centre.

A complete calculation of the quark condensate in the lowest nonvanishing order 
over fluctuations requires solution of the
equations 
\begin{eqnarray}
&&(i\!\not\!D-m)S(x,y)=-\delta(x,y),
\label{quarkprop} \\
&&i\!\not\!\eta(x)e^{i\alpha\gamma_5}S(x,y)=-S(x,y), \ (x-z)^2=R^2,\\
&&S(x,y)i\!\not\!\eta(y)e^{-i\alpha\gamma_5}=S(x,y), \ (y-z)^2=R^2,
\end{eqnarray}
where $\eta_\mu(x)=(x-z)_\mu/|x-z|$, and $D_\mu$ is 
the covariant derivative in 
the fundamental representation,
\begin{eqnarray}
D_\mu=\partial_\mu-i\hat B_\mu
=\partial_\mu + \frac{i}{2}\hat n B_{\mu\nu}x_\nu.
\nonumber
\end{eqnarray}
Substituting 
\begin{equation}
S=(i\!\not\!D+m)
[P_\pm{\cal H}_0+P_\mp O_+{\cal H}_{+1} + P_\mp O_-{\cal H}_{-1}]
\label{q-pr1},
\end{equation}
into Eq.~(\ref{quarkprop}) where
\begin{eqnarray}
O_{\pm} & = & N_+\Sigma_{\pm} + N_-\Sigma_{\mp}, 
\nonumber\\ 
N_\pm & = &  \frac{1}{2}(1\pm \hat n/|\hat n|), 
\nonumber\\ 
\Sigma_\pm & = & \frac{1}{2}(1\pm \vec\Sigma\vec B/B), 
\nonumber
\end{eqnarray}
and $\hat B=|\hat n|B$, shows that the scalar functions
 ${\cal H}_{\zeta}$, with $\zeta = 0,\pm 1$, should satisfy the equations:
\begin{eqnarray}
(-D^2+m^2+2\zeta \hat B){\cal H}_{\zeta}=\delta(x,y).
\end{eqnarray}
We note that if solutions vanishing at infinity were sought,
then the Green function ${\cal H}_{-1}$ would be divergent 
in the massless limit due to the contribution 
of zero modes of the Dirac operator in the presence of the (anti-)self-dual
homogeneous field. The present bag-like boundary conditions remove zero
eigen-values from the spectrum, and the massless limit is regular.
Due to averaging over self- and anti-self-dual configurations and 
all possible values of angle $\alpha$ in the partition function 
chiral symmetry is not broken explicitly.
However, as we show below, a nonzero quark condensate arises 
in the massless limit 
due to an interplay of random distribution of the domains with
self- and anti-self-dual field and the boundary conditions
with the random value of the chirality violating angle $\alpha$.

In order to avoid cumbersome calculations and expose the
role of the former zero modes 
in a transparent way we turn to the particular 
choice $y=z=0$ and calculate the value of the quark condensate at the centre
of the domain.  In this case the functions ${\cal H}_{\zeta}$ 
can depend only on $x_\mu$, $B_{\mu\nu}x_\nu$ and $\eta_\mu=x_\mu/\sqrt{x^2}$,
and hence are functions of $x^2$ only, and the general solutions 
for scalar Green's functions take the form 
\begin{eqnarray}
{\cal H}_{\zeta}=\Delta(x^2|\mu_{\zeta})+
C_{\zeta}\Phi(x^2|\mu_{\zeta}),
\nonumber
\end{eqnarray}
where $\mu_{\zeta}=m^2/2B+\zeta$, and  
\begin{eqnarray} 
\Phi(x^2|\mu)=e^{-Bx^2/4}M(1+\mu,2,Bx^2/2).
\nonumber
\end{eqnarray}
Here $\Delta(x^2|\mu)$ is the vanishing at infinity scalar propagator
with mass $2B\mu$ in the homogeneous (anti-)self-dual field
and $\Phi$ is a solution to the homogeneous equation regular at $x^2=0$,
expressed in terms of the confluent hypergeometric function.
The constants $C_{\zeta}$ can be used to fit the boundary condition.
Terms with ${\cal H}_0$ and ${\cal H}_{+1}$ are regular in the massless limit
and cannot contribute to the trace of the quark propagator.
Thus we concentrate on the term ${\cal H}_{-1}$. Using identities 
\begin{eqnarray}
\gamma_\mu B_{\mu\rho}x_\rho P_\mp\Sigma_+ & = & 
iB\!\not\!x P_\mp\Sigma_+,
\nonumber \\
\gamma_\mu B_{\mu\rho}x_\rho P_\mp\Sigma_- & = &  -iB\!\not\!x P_\mp\Sigma_-,
\nonumber
\end{eqnarray}
one can show that the boundary condition is satisfied if
on the boundary
\begin{eqnarray}
&&2e^{\mp i\alpha}m {\cal H}_{-1} = -2{\cal H}_{-1}' - \hat BR^2{\cal H}_{-1}.
\nonumber
\end{eqnarray} 
which implies that in the massless limit $C_{-1}$ takes the form
\begin{eqnarray}
&&C_{-1}=-\frac{\hat B^2}{4\pi^2m^2}+
\frac{e^{\pm i\alpha}}{2\pi^2R^3m}F(\hat BR^2/2) + O(1),
\nonumber\\
&&F(z)=e^z-z-1+\frac{z^2}{4}\int_0^\infty
\frac{dte^{2t-z({\rm coth}t-1)/2}}{{\rm sinh}^2t} 
({\rm coth}t-1),
\nonumber
\end{eqnarray}
Moreover the singular terms cancel in $\!\not\!D{\cal H}_{-1}$ 
and 
\begin{equation}
\label{zermod}
\lim_{m\to 0}m{\cal H}_{-1}(x,0)=
\frac{e^{\pm i\alpha}}{2\pi^2R^3}F(\hat BR^2/2)e^{-\hat Bx^2/4},
\end{equation}
and thus,
\begin{eqnarray} 
{\rm Tr}S(0,0)=\frac{e^{\pm i\alpha}}{2\pi^2R^3}\sum\limits_{|\hat n|}
F(\hat BR^2/2).
\label{trS}
\end{eqnarray}
It should be noted that the part of the propagator responsible for the
nonzero trace, Eq.~(\ref{zermod}), is proportional to the zero mode of 
the Dirac operator
\begin{eqnarray}
i\!\not\!DP_{\mp}O_- e^{-\hat Bx^2/4}\equiv 0.
\nonumber
\end{eqnarray} 

Now we have to average Eq.(\ref{trS}) over domain configurations taking into 
account the quark determinant. 
According to \cite{wipf} the $\alpha$-dependence of
the quark determinant is  
$$
\exp\left\{ 
\frac{i\alpha}{32\pi^2}\int dx \tilde  B_{\mu\nu}B_{\mu\nu}\theta(1-x^2/R^2)
\right\}=\exp\{\pm i q\alpha\}
$$
where $q$ is the topological charge associated with a domain,
and thus a $\theta$-term is generated effectively by the
quark determinant. 
After averaging over $\alpha$ we get a finite value 
for the condensate at the center of the domain 
\begin{eqnarray}
\langle \bar\psi \psi\rangle= 
-\frac{q}{2\pi^2R^3(1+q)}\sum\limits_{|\hat n|}
F(\hat BR^2/2).
\nonumber
\end{eqnarray}
Numerically this is equal to
$
\langle \bar\psi \psi\rangle= -(228 {\rm MeV})^3
$
for  $B$ and $R$  fixed by the string constant as in 
in Eq.~(\ref{par-val}).

\section{Propagators in the Presence of Domains}

In order to study in more detail the influence of domain structure 
and the mean field on the properties of the dynamical quarks and gluons
we have to find their propagators. 
They
can be analytically calculated by reduction 
to the scalar problem, essentially that  
of a four-dimensional harmonic oscillator with  
total angular momentum coupled to the external field. 
The general solution is given by decomposition over hyperspherical harmonics.
In the following section we present the exact solution for the 
scalar propagator, though with most derivations
relegated to Appendices D and E. With the scalar result we derive propagators 
for ghost and gluon fluctuations in an external (anti-)self-dual field 
with Dirichlet boundary conditions imposed on the fluctuations on
a hyperspherical surface.

\subsection{Scalar Propagator}

The problem to be solved is given by the scalar Green's function equation  
\begin{eqnarray}
&&[ (\partial_{\mu} - i B_{\mu})^2 -{\cal M}^2] G(x,x'|\mu) 
= - \delta^{(4)}(x-x'),
\nonumber\\
&&B_{\mu}  =  - {\frac{1}{2}}B_{\mu \nu} x_{\nu}, \
B_{\mu \nu} B_{\mu \rho}  =  B^2 \delta_{\nu \rho} ,
\label{bgconv}
\end{eqnarray}
with the homogeneous Dirichlet boundary condition
$$
G(x,x'|\mu)_{x^2=R^2} = G(x,x^\prime|\mu)_{x^{\prime 2}=R^2} = 0,
$$
where $R$ is the radius of a hypersphere centred
at the origin and 
$\mu={\cal M}^2/2B$.

We present first the solution to the corresponding eigenvalue problem,
\begin{eqnarray} 
- [ (\partial_{\mu} - i B_{\mu})^2 -{\cal M}^2]
{\psi}_{\lambda} = \lambda {\psi}_{\lambda}.
\nonumber
\end{eqnarray}
One may choose $B_{\mu \nu}$ such that
$$
B_{34} = E, B_{12} = B, -B \leq E \leq B.
$$
A representation of the eigenfunctions in terms of a complete
orthonormalised set of eigenfunctions of the four-dimensional
Laplace operator is achieved
in the following hyperspherical coordinate system
(see for example, \cite{Sto56})
\begin{eqnarray}
x_1 & = & r \sin \eta \cos \phi \nonumber \\
x_2 & = & r \sin \eta \sin \phi \nonumber \\
x_3 & = & r \cos \eta \cos \chi \nonumber \\
x_4 & = & r \cos \eta \sin \chi .
\end{eqnarray}
The angular eigenfunctions are 
\begin{eqnarray}
C_{k m_1 m_2}(\eta,\phi,\chi) & = &
(-1)^{|m_1 + m_2|} (2\pi)^{-1} \Theta_k^{m_1-m_2,m_1+m_2}(\eta),
\nonumber \\
{}&{}& \times \exp i[(m_1-m_2)\chi + (m_1 +m_2)\phi] \nonumber \\
\Theta_k^{k-r-s,s-r}(\eta) & = & \sqrt{2(k+1)(k-r)!(k-s)! r! s!}
\nonumber \\ 
{}&{}& \times \sum_{n=0}^r 
{{ (-1)^{r-n} \cos^{k-r-s+2n} \eta \sin^{r+s-2n}\eta } 
\over
{(k-r-s+n)! n! (r-n)!(s-n)!} } \nonumber \\
{} &{}& \quad \quad  (r,s=0,1,\dots,k),
\nonumber
\end{eqnarray}
where $k, m_1, m_2$ are respectively the orbital angular
momentum and the two azimuthal quantum numbers, relevant for
a four-dimensional hyperspherical symmetry.
That the $C_{k m_1 m_2}$ are eigenfunctions with the said
eigenvalues is proven in Appendix D.

The eigenfunctions for the complete problem are a product of radial and
angular parts,
$$
\psi(x) = f(r) C_{k m_1 m_2}(\eta,\phi,\chi).
$$
The radial equation has a solution expressed in terms of the confluent
hypergeometric function, in the notation of \cite{AS65}, 
\begin{equation}
f(r) = (Br^2/2)^{k/2}   
e^{-Br^2/4}
M\left(\frac{k}{2} + 1 - m_{2,1} + {{{\cal M}^2-{\lambda}} \over {2B}} , k+2; Br^2/2\right),
\label{radialsol}
\end{equation}
where the function regular at $r=0$ is chosen for normalisability.
Here $m_{2,1}$  should be put equal to  $m_2$  for the self-dual field,  and  
$m_1$ for the anti-self-dual field.  It is convenient to denote
\begin{equation}
n_{2,1} = \frac{k}{2} + 1 - m_{2,1}.
\label{index}
\end{equation}
Another independent solution (not normalisable in our problem), which is 
regular at infinity and singular at the origin,  would 
be obtained by replacing the function $M$ with the function $U$.
Imposition of the Dirichlet condition at $r=R$ forces
eigenvalues $\lambda$ to take discrete values defined by the zeroes of 
$M(a,b,z)$ as a function of $a$ at fixed $b$ and $z$. The eigenvalues 
$\lambda$ are strictly positive.
As we will see below, the case of ${\cal M}^2=-2B$ will be met in the problem 
for gluon propagator.
In this case the lowest eigenvalue $\lambda_0$ is defined by ($k=m_1=m_2=0$)  
$$
M\left(-\lambda_0 , 2; BR^2/2\right)=0,
$$
and is a positive function of $BR^2$, as can be checked.

The propagator can be found by the standard method of decomposition over 
hyperspherical harmonics, see for instance~\cite{MF53}, 
exploiting the above hyperspherical 
representation. The derivation is given in Appendix E 
but it consists of essentially two steps. First we exploit 
completeness of the angular eigenfunctions in order
to reduce the four dimensional problem to a one-dimensional
Sturm-Liouville problem with known exact solution for the radial
dependence. Second we add a solution to the homogeneous
equation with coefficient selected to implement the above
boundary condition.  

Using the solution to the eigenvalue equation with $\lambda=0$, 
one gets independent solutions for the homogeneous equation. 
In the notation of \cite{AS65}, 
the two solutions (respectively regular at the origin and at infinity) are
\begin{eqnarray}
\label{r1r2}
R_1(r|k,n_{2,1},\mu) & = & 
r^k e^{-Br^2/4} M(n_{2,1} + \mu,
k+2; Br^2/2) \nonumber \\
R_2(r|k,n_{2,1},\mu) & = & 
r^k e^{-Br^2/4} U(n_{2,1} + \mu,
k+2; Br^2/2) . 
\end{eqnarray}
The Sturm-Liouville equation is satisfied by  
\begin{eqnarray}
X_{k n_{2,1}}(r,r'|\mu) = B
{{\Gamma(n_{2,1} + \mu)}  \over { 4 \Gamma(k+2) } } 
R_{k n_{2,1}}(r,r'|\mu)
\end{eqnarray}
with 
\begin{equation}
\label{r12}
R_{k n_{2,1}}(r,r'|\mu) = \left\{ 
\begin{array}{c}   R_1(r|k,n_{2,1},\mu) R_2(r'|k,n_{2,1},\mu), \ r < r'  \\
                   R_1(r^\prime|k,n_{2,1},\mu) R_2(r|k,n_{2,1},\mu), \ r > r'
\end{array} \right. .
\end{equation}
In terms of these quantities, the Green's function is:  
\begin{eqnarray}
\label{fin-G1}
&&G(x,x'|\mu) =  \frac{B}{4}\sum_{k,m_1,m_2} 
{{\Gamma(n_{2,1} + \mu)}  \over { \Gamma(k+2) } } 
C_{k m_1 m_2}(\eta',\phi',\chi')
C_{k m_1 m_2}(\eta,\phi,\chi) 
\\
&&\times \left[ R_{k n_{2,1}}(r,r'|\mu)  
 -  { {U(n_{2,1} + \mu, k+2;BR^2/2)}
\over
  {M(n_{2,1} + \mu, k+2;BR^2/2)} } 
R_1(r|k, n_{2,1},\mu) R_1(r'|k, n_{2,1},\mu)
\right].
\nonumber
\end{eqnarray}
The first term inside square brackets guarantees $G$ to be a Green's function 
through the solution to the Sturm-Liouville equation. 
The coefficient in the second term is determined by the boundary condition. 
The only singularity in $G$ is the usual ultraviolet one at $x=x^\prime$. 
For more details we refer the reader to Appendix~D.

\subsection{Ghost and Gluon Propagators}

We work in the background gauge 
$$
D_\mu^{ab}A_\mu^b=0,
$$
and use conventions for the adjoint representation of colour $SU(3)$ 
\begin{eqnarray}
&&{\breve D}_\mu=\partial_\mu-i\breve n B_\mu, \ \breve n=T^an^a,
 \ T^a_{bc}=-if^{abc}.
\nonumber\\
&&\breve n =T^3\cos\xi_k +T^8\sin\xi_k , \ \xi_k=(2k+1)\pi/6, \ k=0,1,\dots,5.
\label{xivals} \\
&&\breve B_{\mu\nu}=\breve n B_{\mu\nu}, \
\breve B= \sqrt{\breve n^2} B.
\nonumber
\end{eqnarray}
The colour matrix 
$\breve{n}$ can be diagonalised by the unitary transformation
 \begin{eqnarray}
&&n=U \breve{n} U^{\dagger} 
= {\rm diag}\left( 
 \zeta_1, -\zeta_1,0,  \zeta_2, -\zeta_2 ,  \zeta_3, -\zeta_3, 0 \right)
\nonumber\\
&&\zeta_1=\sin \xi,   \
\zeta_2=(\sin\xi +\sqrt{3}\cos\xi)/2 , \
\zeta_3=(-\sin\xi +\sqrt{3} \cos\xi)/2,
\nonumber
\end{eqnarray}
where
$U 
={\rm diag}(W,0,W,W,0)$ with 
\begin{equation}
W
={1\over \sqrt{2}} 
 \left( \begin{array}{cc}
 1   & -i  \\
   1 & i  \\
\end{array} \right).
\end{equation}
For the values of the
angle $\xi_k$ as in Eq.(\ref{xivals}) the $\zeta_i$ take values from the set
$(\pm 1, \pm 1/2)$. Namely, for $k=0,1,2,3,4,5$ respectively,
the three $\zeta_i$ are
$(1/2,1,1/2)$, $(1,1/2,-1/2)$, $(1/2,-1/2,-1)$, $(-1,-1/2,1/2)$,  
$(-1/2,1/2,1)$.
The diagonalized covariant derivative takes the form
\begin{equation}
D_\mu =U \breve D_{\mu} U^{\dagger}   
= \partial_{\mu} -i n B_{\mu}.
\end{equation}
In the Feynman gauge, diagonalized equations for the ghost and gluon  
propagators take the form  
 \begin{eqnarray}
&& -D^2G(x,x') = \delta(x-x'),
\nonumber\\
&&
\label{vect-1}
\left(-D^2\delta_{\mu\nu}+2in B_{\mu\nu}
\right)G_{\nu\rho}(x,x')=\delta^{\mu\rho}\delta(x-x'),
\end{eqnarray} 
with the diagonalized boundary conditions
\begin{eqnarray}
nG_{\nu\rho}(x,x')=0 ,  \ n G(x,x')=0 , \  {\rm {for}} \  x^2=R^2  \
{\rm {or}} \  x^{\prime 2}=R^2,
\nonumber
\end{eqnarray}
where $n$ and the propagators are diagonal matrices in colour indices. 
As evident above, the matrix $n$ has two zero eigenvalues and the 
corresponding gluon components 
are not restricted by above boundary condition so the equations for these 
modes are simply the free ones.
These modes are thus not confined in the model under consideration. 

The scalar equation for the ghost propagator has been solved in the previous 
section, where one should simply replace  $B\to nB$ and put ${\cal M}=0$.
The equation for the gluon propagator and the boundary condition can be 
further diagonalised with respect to Lorentz indices
and is thus reduced to four scalar equations, each of which has a well defined 
solution since, as discussed above,
dangerous zero modes, so called chromons~\cite{leutw,mink},
do not satisfy Dirichlet conditions and do not contribute to the Green's
function.  The original propagators are restored by the inverse transformations.

In the case of the gluon propagator  
one can avoid the second diagonalization by looking for a solution of the form 
\begin{eqnarray}
\label{vecsol-1}
G_{\nu\rho}(x,x')=\left(
D^2\delta_{\nu\rho}+2in B_{\nu\rho}
\right)\Delta(x,x').
\end{eqnarray} 
Substitution of Eq.~(\ref{vecsol-1}) into the original Green's function
equation gives
\begin{eqnarray}
\left[
-(D^2)^2+4n^2 B^2
\right]\Delta(x,x')=\delta(x-x'),
\nonumber
\end{eqnarray} 
with the solution given formally by
\begin{eqnarray}
\Delta(x,x') 
& = & {1 \over {-(D^2)^2 + 4 n^2 B^2} }\delta(x-x') 
\nonumber \\
& = & {1 \over {4 |n| B}} 
    \left( {{1} \over {-D^2 +2 |n| B)}}
          - {1 \over {-D^2 - 2 |n| B}} \right) \delta(x-x').
\label{del} 
\end{eqnarray} 
The two terms above are nothing but scalar propagators
with ``mass term''  ${\cal  M}^2=\pm 2|n| B$.
Substituting Eq.~(\ref{del}) into Eq.~(\ref{vecsol-1}) and using notation 
of the previous subsection 
for the scalar propagator with $B\to |n|B$ and $\mu=\pm 1$ one gets after 
simple manipulations
\begin{eqnarray}
\label{vecsol-f}
G_{\mu\nu}(x,x') = {1 \over {2}}\delta_{\mu\nu}
\left[
G(x,x'|1)+G(x,x'|-1)
\right] 
+\frac{inB_{\mu\nu}}{2|n|B}
\left[
G(x,x'|1)-G(x,x'|-1)
\right] ,
\end{eqnarray}
where only nonzero elements of the diagonal matrix $n$  are involved.
With this representation it is clear that the boundary condition for 
the gluon propagator is satisfied if the scalar Green's functions 
$G(x,x'|\pm1)$ 
are subject to the homogeneous Dirichlet condition independently of each other.
An explicit form is obtained via Eq.~(\ref{fin-G1}) by substitution  
$B\to|n|B$  and $\mu=\pm1$.
This can be done straighforwardly for the terms with $k-m_{2,1}>0$ 
in the expansion over hyperspherical harmonics, as is obvious from the integral 
representations of the confluent hypergeometric functions~\cite{AS65}
($b>a>0$)
\begin{eqnarray}
\label{conf-rep}
&&M(a,b,z)=\frac{\Gamma(b)}{\Gamma(a)\Gamma(b-a)}\int_0^1dte^{zt}
t^{a-1}(1-t)^{b-a-1},
\nonumber\\
&&U(a,b,z)=\frac{1}{\Gamma(a)}\int_0^\infty dt e^{-zt}t^{a-1}(1+t)^{b-a-1},
\end{eqnarray} 
with $a=1+\mu+k/2-m_{2,1}$ and $b=k+2$, in our case.
Special comment is required for the terms with 
$a=k/2-m_{2,1}=0$ in the decomposition of $G(x,x'|-1)$.

Using the representations Eqs.(\ref{conf-rep}) it immediately follows that 
$$
\lim\limits_{a\to0}M(a,b,z)=1+O(a),  \ \lim\limits_{a\to0}U(a,b,z)=1+O(a).
$$
With this and Eqs.~(\ref{r1r2})-(\ref{fin-G1}) one can be convinced that 
the singularity 
in the gamma-function in Eq.~(\ref{fin-G1}) at $a=n_{2,1}-1=k/2-m_{2,1}=0$ 
is cancelled by the contribution coming from the expression in the square 
brackets.  Thus we conclude that the gluon propagator exists. 
This confirms the absence of the zero modes under the imposed Dirichlet 
boundary conditions. 
 
\subsection{A comment on analytical properties}

The scalar propagator Eq.(\ref{fin-G1})
which determines the analytic properties
of the off-diagonal  components of the ghost and gluon propagators
have compact support in hyperspherical region of radius $R$ in Euclidean space-time
with the usual ultraviolet integrable singularity at $x'=x$. 
Thus the Fourier transform of the propagator 
averaged over domain position, given by the integral
\begin{eqnarray}
\tilde G(p^2)=\int_{V_{R}} d^4x e^{ipx} G(x),
\nonumber\\
G(x-y)=v^{-1}\int_V dz G(x-z,y-z), 
\end{eqnarray}
leads to a $\tilde G(p^2)$ which is an entire analytical function 
in the complex $p^2$ plane. 
Entire propagators are typical for nonlocal field theories and 
has been interpreted as confinement of  
dynamical charged fields \cite{leutw,efiv,efned}. 
Thus the presence of domains
maintains confinement of off-diagonal gluons and ghosts.

An instructive example is given by a toy calculation which illustrates 
the qualitative behaviour of the Fourier transform of propagators 
with compact support in a finite region of ${\it R}^4$.
We calculate the Fourier transform of the function
$$
D(x)=\frac{\theta(1-x^2/R^2)}{4\pi^2x^2}.
$$
The calculation proceeds via the following steps (where $p=|p|$)
\begin{eqnarray}
\tilde D(p)&=&\frac{1}{\pi}\int_{-1}^1dt\sqrt{1-t^2}\int_0^Rdrre^{iprt}
=\frac{2}{\pi}\int_{0}^1dt\sqrt{1-t^2}\int_0^Rdrr \cos(prt]
\nonumber\\
&=&\frac{1}{p}\int_0^RdrJ_1(pr)=\frac{1}{p^2}\int_0^{Rp}dxJ_1(x)
=\frac{2}{p^2}\sum\limits_{k=0}^{\infty}J_{2k+2}(Rp)
\nonumber\\
&=&\frac{1-J_0(Rp)}{p^2},
\end{eqnarray}
where we have used the identity
$$
J_0(z)+2\sum\limits_{k=1}^{\infty}J_{2k}(z)=1.
$$
This propagator is an entire function with the properties
\begin{eqnarray}
&&\tilde D(0)=R^2/4, \ \tilde D(ip)=\frac{I_0(Rp)-1}{p^2}, 
\nonumber\\ 
&&\lim_{p^2\to\infty}\tilde D(p)=
\frac{1}{p^2}\left[1-\sqrt{\frac{2}{\pi Rp}}\cos(Rp-\pi/4)\right],
\nonumber\\
&&
\lim_{p^2\to\infty}\tilde D(ip)=
\frac{e^{Rp}}{p^2\sqrt{2\pi Rp}},
\end{eqnarray}
which indicate the standard $p^{-2}$ 
behaviour for asymptotically large Euclidean momenta, 
and an exponential rising in the physical region (large energy).
We intend to consider elsewhere detailed analytic properties of 
the Green's functions in the present approach (including that for fermions).

\section{Conclusions and Open Problems}

The idea of domains in the vacuum is not a new one and various hints and 
attempts at implementation of such an idea 
can be found \cite{leutw,sim,Amb80,AmS90}.
These approaches assume explicitly or implicitly that the boundaries of domains 
are populated by the (chromo)electric and/or (chromo)magnetic ``charges
and/or currents'' which produce nonzero field strength inside domains.
Thus the source for the mean field inside is assumed to be present on the 
boundary.  Specific configurations suitable in 
principle for a description of such domains are known
(see for instance \cite{sim,toron,Zhit90,dyon,chak}).
In this picture domains are assumed to be stable 
and in this sense are somewhat similar to the usual domains in
ferromagnets.
  
The model presented in this work differs cardinally from this picture. 
The central idea that enables us to introduce and consider domains
is the observation made in Refs.~\cite{lenz1,lenz2}
that the presence of singular pure gauge background fields
imposes specific conditions on quark, ghost and gluon fluctuations. 
The boundaries correspond to the locations of singularities 
in the pure gauge vector potentials which by themselves do not generate  
any field strength.  Such boundaries  make their presence felt 
only via their impact on quantum fluctuations. The mean field inside
domains appears as a collective effect of quantum fluctuations,
 which themselves remain subject to certain boundary conditions.
The domains are not stable in this picture, but describe
a specific class of field fluctuations in the system. 
Within this model all the fundamental features of the QCD vacuum 
-- gluon condensation, topological susceptibility, confinement 
of static and dynamical charges and a non-zero quark condensate --
emerge in a transparent and simple way.

So far we have discussed this mechanism in a purely qualitative manner and
the relationship of the model with real QCD has to be 
clarified.  It should be recalled that our motivation skipped over  
two points, both requiring more formal justification.
In the first step we prescribed a particular way of dealing with
singular pure gauges and thus the QCD functional integral 
incorporated densely packed interacting domains.
In the second step we replaced this integral by a model partition function
describing decoupled hyperspherical domains or clusters.
The role of the mean field inside domains is to compensate effectively for 
the decoupling.
The problem of verification of both steps remains open.
In particular, a possible relationship between this kind of domain formation 
via singular pure gauges and the Gribov problem has yet to be understood.  
 
Concerning phenomenological applications,
a complete solution to the fermionic eigenvalue problem would immediately
enable clarification of the connection between the
picture of spontaneous chiral symmetry breaking in this model
and the  Banks-Casher relation \cite{BC80}. 

With quark and gluon propagators in the mean field
detailed applications to hadron physics are accessible. 
The meson spectrum, for example, can be computed via
a bosonisation procedure as applied in \cite{efned,bur}  or via Bethe-Salpeter
equations. 
Entire quark and gluon propagators are expected to give rise to the Regge character of the spectrum of 
relativistic bound states~\cite{bur,Efi01}. 

In this context the $U_A(1)$ problem can be also addressed.
Preliminary estimations show that due to nonzero topological susceptibility the 
pseudoscalar correlators in the isovector and isoscalar channels 
are different in the massless limit and strong splitting 
between the masses of the $\eta'$ and $\pi$ mesons is expected,
Alternately,  the
anomalous Ward identity of \cite{Cre77} could be studied order by order 
in the decomposition over fluctuation fields.
The fact that for the pure glue theory a reasonable value
for the topological susceptibility is obtained simultaneous with a non-zero
quark condensate is encouraging in this respect.

It would be tempting to look for the present picture in
lattice simulations. However,  
domains of constant field can only be taken seriously in
a statistical sense, so one should compare results not configuration
by configuration (say, after moderate cooling)  but for correlators and
condensates calculated within the model and on the lattice, where a
full statistical ensemble has been taken into account.

Returning to gluonic fluctuations, the picture of dynamical
confinement remains incomplete. Diagonal or ``neutral'' gluons
remain freely propagating modes in the pre-mean field framework. 
Intuitively it is clear that this problem is ultimately related
to the topological triviality of the class of
singular field we have considered here. Incorporating a wider hierarchy
of singular fields can resolve this problem.

Finally, a set of open problems relate to the general
properties of quantum field theory with
domain-like structures and Dirichlet boundary conditions 
on fluctuation fields. Entire propagators,  
which appear as a result, indicate that the theory is nonlocal.
The ramifications of non-locality need to be investigated,
particularly in light of recent work by Efimov \cite{Efi97}.
Although the choice of boundary conditions is not expected  
to generally influence short-distance singularities in Green's functions,
the explicit structure of ultraviolet divergencies and the question of 
renormalisability of a quantum field model with 
Dirichlet boundary conditions imposed on fields 
in regions of space should be investigated explicitly.

\acknowledgements
ACK is supported under a grant from the Australian Research Council
and in the initial phase of this work under the grant 06 ER 809 of
the BMBF.  SNN was partially supported by the grant RFFI 01-02-17200.  
The authors are grateful to Frieder Lenz for 
fruitful discussions, critical
comments and suggestions.   Garii Efimov,   Andreas Schreiber,
Gerald Dunne, Jan Pawlowskii and Lorentz von Smekal   
are also thanked for numerous constructive discussions,
as well as Max Lohe for tips in dealing with the hyperspherical 
eigenfunctions.  

\appendix

\section{ Quantum Electrodynamics}

The purpose of the following is to illustrate how the effective action
as a functional of the mean field and characteristic functions can be defined 
formally in the abelian case. 
Consider QED 
\begin{eqnarray}
L=-\frac{1}{4}Q_{\mu\nu}Q_{\mu\nu}+\bar\psi(i\!\not\!\partial-m-e\!\not\!Q)\psi
-e\bar\psi\!\not\!S\psi
\end{eqnarray}
in the presence of an external pure gauge singular field of the form
$$
S_\mu=\sum\limits_j^N \partial_\mu f_j(x),
$$ 
where the functions $f_j$ have topologically trivial singularities on 
hypersurfaces $\partial V_j$
and are assumed to be not Fourier transformable. Gauge transformations
which would remove such a pure gauge field are then not defined.  
Here $Q_{\mu\nu}=\partial_\mu Q_\nu-\partial_\nu Q_\mu$
is the field strength for the photon  fields.  
Thus $S$ appears only in the interaction term coupling to the fermion field.
The fluctuation fields $Q$, $\psi$ and $\bar \psi$  are assumed to be regular 
differentiable functions everywhere in Euclidean space.
It should  be stressed that unlike non-abelian theory 
there is no internal necessity for considering singular fields 
in electrodynamics, and the example below is artificial in this sense. 

The field $S$
in the vicinity of the $j-$th singular surface can be represented as
\begin{eqnarray}
S_\mu \sim \eta^j_\mu (\eta^j_\nu\partial_\nu)f_j(x), 
\end{eqnarray}
where $\eta^j_\mu$ is a unit vector normal to the surface $\partial V_j$.
Finiteness of the action density thus requires that
\begin{eqnarray}
\bar\psi(x)\!\not\!\eta^j(x)\psi(x)=0, \ x\in\partial V_j.
\end{eqnarray}
This condition is satisfied if, for $x$ on the boundary
\begin{eqnarray}
\label{fbc}
\psi=-i\!\not\!\eta^j e^{i\alpha_j\gamma_5}\psi,
\
\bar\psi=\bar\psi i\!\not\!\eta^j e^{-i\alpha_j\gamma_5},
\end{eqnarray}
which is the well-known bag-like boundary condition \cite{wipf}.
Note that we are working in Euclidean space-time and the fields
$\psi$ and $\bar\psi$ are independent variables. The angle $\alpha_j$
is arbitrary and need not be the same for different $j$.
It should be stressed that the boundary condition violates
chiral symmetry.
No conditions on the photon fluctuation field $Q$ arise since it is decoupled 
from $S$.  Now we can write down the functional integral straightforwardly
\begin{eqnarray}
\label{qed-1}
{\cal Z}[S]=\int{\cal D} Q \delta (\partial Q)
\int_{{\cal F}_S}{\cal D} \psi {\cal D} \bar \psi
e^{-{\cal S}[Q+S,\psi,\bar \psi]}, 
\end{eqnarray} 
where now the space ${\cal F}_S$ contains 
only those fields which satisfy the boundary conditions Eq.(\ref{fbc}).
We stress that the field $S$ in Eq.~(\ref{qed-1}) is considered as a 
fixed background field. Gauge fixing for the field $Q$ can be achieved  
by regular gauge transformations.
We see from Eq.(\ref{qed-1}) that due to the presence of the singular 
field $S$, the integral over fermionic fluctuations
is separated into integrations over fields inside subregions $V_j$
bounded by the surfaces $\partial V_j$ where the background field is singular,
and these fluctuations are subject to the boundary conditions (\ref{fbc}).

Now we define a procedure for averaging over singular 
configurations. This is done by identifying
the set of different singular configurations with a set of 
characteristic functions dividing Euclidean space into subregions, 
whose boundaries coincide with the singular surfaces of a given singular 
field
\begin{eqnarray}
\label{aver-1}
&&\left\{S\right\}\longleftrightarrow 
\left\{\chi_1,\dots,\chi_N\right\},
\
\sum\limits_j^N\int_V d^4x\chi_j^\kappa(x)=V,
\nonumber
\end{eqnarray}
where the requirement of conservation of the total volume is imposed.
Integration over $S$ is defined as an averaging over an ensemble
of characteristic functions and angles $\alpha_j$ coming through the 
fermionic boundary conditions,
\begin{eqnarray}
\label{aver-2}
\int{\cal D} S {\cal Z}[S] & \longleftrightarrow &  
\prod\limits_j^N\int {\cal D}\chi_j d\alpha_j
\delta\left(1-V^{-1}\sum_k^N\int d^4x\chi_k(x)\right)
\nonumber \\
{}&{}& \times {\cal Z}[\chi_1,\dots,\chi_N|\alpha_1,\dots,\alpha_N],
\\
{\cal Z}[\chi|\alpha] & = &\prod\limits_j^N\int{\cal D} Q\delta(\partial Q)
\int\limits_{{\cal F}_j(\alpha_j)}{\cal D}\psi_j{\cal D}\bar\psi_j
\nonumber \\
{}&{}& \times \exp\left\{
-\int_{V_j}d^4x
\left[\frac{1}{4}Q_{\mu\nu}^2-\bar\psi_j(i\!\not\!\partial-m-
e\!\not\!Q)\psi_j
\right]
\right\}.
\nonumber \\ 
\end{eqnarray}
In this representation translation invariance as well as
chiral symmetry (for $m=0$) are restored because of the averaging
over all $\alpha_j$ and characteristic functions $\chi_j$.

Let us integrate out the photon field and factorise the part of 
the fermionic integral corresponding to $k-$th region. We obtain  
\begin{eqnarray}
{\cal Z}_k[\chi|\alpha] &=&  
\int\limits_{{\cal F}_k(\alpha_k)}{\cal D}\psi_k{\cal D}\bar\psi_k
\prod\limits_{j\not=k}^N
\int\limits_{{\cal F}_j(\alpha_j)}{\cal D}\psi_j{\cal D}\bar\psi_j
\nonumber \\
{}&{}& \times \exp\left\{
\int_{V_k}d^4x
\bar\psi_k(i\!\not\!\partial-m)\psi_k 
+ 
\frac{e^2}{2}\int_{V_k}d^4xd^4y J_k(x)D(x-y)J_k(y)
\right\}
\nonumber\\
{}&{}& \times \exp\left\{
\int_{V_j}d^4x
\bar\psi_j(i\!\not\!\partial-m)\psi_j
 +e^2\sum\limits_{j\not=k}
\int_{V_k}d^4x\int_{V_j}d^4y J_k(x)D(x-y)J_{j}(y)
\right. \nonumber \\
{}&{}&
+\left. \frac{e^2}{2}\sum\limits_{j,j'\not=k}
\int_{V_j}d^4x\int_{V_j'}d^4y J_j(x)D(x-y)J_{j'}(y)
\right\},
\nonumber
\end{eqnarray}
where $J_j(x)$ denotes the electromagnetic current and $D(x-y)$ is the 
standard photon propagator. Inserting `unity' represented as  
\begin{eqnarray}
1=\prod_{x\in V_k}\int {\cal D} B^k 
\delta\left[
B^k-e\sum\limits_{j\not=k}\int_{V_j}d^4y D(x-y)J_j(y)
\right],
\nonumber
\end{eqnarray}
we arrive at the representation
\begin{eqnarray}
\label{qed-2}
{\cal Z}[\chi|\alpha] & = &\int {\cal D} B^{k}
\exp\left\{-{\cal S}_{\rm eff}[B^k|\chi,\alpha]\right\}
 \\
&&\times
\int\limits_{{\cal F}_k(\alpha_k)}{\cal D}\psi_k{\cal D}\bar\psi_k
\exp\left\{
\int_{V_k}d^4x
\bar\psi_k(i\!\not\!\partial-m+e\!\not\!B^k)\psi_k \right. \nonumber \\
{}&{}& \left. +  
\frac{e^2}{2}\int_{V_k}d^4xd^4y J_k(x)D(x-y)J_k(y)
\right\},
\nonumber\\
\label{s-eff}
e^{-{\cal S}_{\rm eff}[B^k|\chi,\alpha]}&=&\prod\limits_{j\not=k}^N
\int\limits_{{\cal F}_j(\alpha_j)}{\cal D}\psi_j{\cal D}\bar\psi_j
\delta\left[
B^k-e\sum\limits_{j\not=k}\int_{V_j}d^4y D(x-y)J_j(y)
\right]
\nonumber\\
&&\times\exp\left\{
\int_{V_j}d^4x
\bar\psi_j(i\!\not\!\partial-m)\psi_j
\right. \nonumber \\
{}&{}& \left. +\frac{e^2}{2}\sum\limits_{j,j'\not=k}
\int_{V_j}d^4x\int_{V_j'}d^4y J_j(x)D(x-y)J_{j'}(y)
\right\}.
\end{eqnarray}
In this representation the partition function ${\cal Z}[\chi|\alpha]$
is defined by the fluctuations of the fermion field in an arbitrarily chosen
subregion $V_k$ in the presence of the electromagnetic field $B_\mu^k$,
the dynamics of which are governed by the effective action $S_{\rm eff}$.
 
As is seen from Eq.~(\ref{s-eff}), this effective action has appeared as 
an integral (or collective effect) of 
field fluctuations in the rest of infinite system, outside the $k-$th domain.
This action functionally depends on a division
provided by a particular set of characteristic functions.

Physically different situations would arise depending on the
properties of the effective action. 
This becomes obvious if we write down the partition function 
averaged over an ensemble of the characteristic functions
and boundary conditions (angles $\alpha_j$),
\begin{eqnarray}
\label{qed-3}
{\cal Z} &=& 
\prod\limits_j^N\int {\cal D}\chi_j d\alpha_j
\delta\left(1-V^{-1}\sum_i^N\int d^4x\chi_i(x)\right)
\int {\cal D} B^{k}
\exp\left\{-{\cal S}_{\rm eff}[B^k|\chi,\alpha]\right\}
\nonumber \\
&&\times
\int\limits_{{\cal F}_k(\alpha_k)}{\cal D}\psi_k{\cal D}\bar\psi_k
\exp\left\{
\int_{V_k}d^4x
\bar\psi_k(i\!\not\!\partial-m+e\!\not\!B^k)\psi_k \right.\nonumber \\
{}&{}& \left. + 
\frac{e^2}{2}\int_{V_k}d^4xd^4y J_k(x)D(x-y)J_k(y)
\right\}.
\end{eqnarray}
Two qualitatively different pictures are possible.
If the functional $S_{\rm eff}$ has an absolute minimum at $B_k=0$
and infinitely small volumes of all regions $V_j$ ($j\not=k$),
then we recover the standard QED partition function in the infinite volume.
If, however, the minimum is at nonzero mean field and some nonvanishing
averaged size of subregions is supportable, we depart from standard 
electrodynamics.
In principle the effective action can be calculated (at least within 
perturbation theory). As mentioned in the
introduction, there is no reason to expect that the second scenario
is realised in electrodynamics.

\section{Effective potential for the constant  field}
\

Consider a covariantly constant abelian field with the field strength
parametrised as
\begin{eqnarray}
&&B^a_{\mu\nu}=n^aB_{\mu\nu}, \ 
\ n^aT^a=T^3\cos\xi+T^8\sin\xi,\  
\nonumber\\
&&E_i=B_{4i}, \ H_i=\frac{1}{2}\epsilon_{ijk}B_{jk},\
{\bf E}^2+{\bf H}^2=2B^2, 
\nonumber\\
&&{\bf EH}=|{\bf E}||{\bf H}|\cos\omega, 
\ ({\bf EH})^2={\bf H}^2(2B^2-{\bf H}^2)\cos^2\omega,
\nonumber
\end{eqnarray}
and let the gauge invariant effective potential be given by the series
\begin{eqnarray} 
U_{\rm eff}(B,\omega,\xi)=\sum\limits_{k=1}^\infty 
A_k
{\rm Tr}\breve B^{2k}, 
\ \
(\breve B^{2k})_{\mu\nu}=
\breve n^{2k}B_{\mu\alpha_1}\dots B_{\alpha_{k-1}\nu},
\end{eqnarray}
with $A_k$  constants.  One can show that  if $U_{\rm eff}$
is bounded from below and has a nontrivial minimum as 
a function of parameter $B$
then there is a set of twelve discrete minima corresponding to
an (anti-)self-dual fields and six values of the angle $\xi$. 

The odd powers of $\breve n$ and $B$ do not appear
in the potential, since this would mean violation of Weyl symmetry
and parity respectively. The Weyl group is a discrete subgroup of  
global $SU(3)$ and in this 
case can be seen as the group of permutations of the eigenvalues 
of the matrix $\breve n$. 
Such permutations can be arranged by a shift of the
angle parametrising the abelian field configuration, $\xi\to\xi+\pi n/3$.
In other words, the effective potential is periodic in $\xi$
with a period $\pi/3$, the angle of $SU(3)$.  It is also periodic in $\omega$
with the period $\pi$ due to invariance under parity. 
This can be checked using formulae 
\begin{eqnarray}
{\rm Tr}\breve n^{2}&=&3, \ \ {\rm Tr}\breve n^4=(9/4), \ \ 
{\rm Tr}\breve n^6=(3/16)[10+\cos(6\xi)],
\nonumber\\
\label{tr2}
{\rm Tr}B^2&=&-2({\bf E}^2+{\bf H}^2)=-4B^2, 
\nonumber\\ 
{\rm Tr}B^4&=&2[({\bf E}^2+{\bf H}^2)^2-2({\bf EH})^2]=
8\left[B^4-\frac{1}{2}({\bf EH})^2\right], 
\nonumber\\ 
{\rm Tr}B^6&=&-2({\bf E}^2
+{\bf H}^2)[({\bf E}^2+{\bf H}^2)^2-3({\bf EH})^2]\nonumber\\
&=&
-16B^2\left[B^4-\frac{3}{4}({\bf EH})^2\right].
\nonumber
\end{eqnarray}
A nontrivial dependence on $\xi$ appears for $k\ge3$. 
Higher terms depend on $\xi$ via functions  $\cos6l\xi$ ($l\ge 1$).
Taking into acount the first three terms in the decomposition
and calculating the traces as above one gets
\begin{eqnarray}
U_{\rm eff}
=-C_1B^2
+\frac{C_2}{\Lambda^4}\left[B^4-\frac{1}{2}({\bf EH})^2\right]
+\frac{C_3}{\Lambda^8}B^2(10 +\cos6\xi)
\left[B^4-\frac{3}{4}({\bf EH})^2\right].
\nonumber 
\end{eqnarray}
The coefficients $C_2$ and $C_3$ are assumed to be positive to provide for 
the boundedness of the potential from below, and $\Lambda$ is a scale. 
The sign of the constant $C_1$ is of particular importance.
If $C_1$ is negative the minimum
is trivial  $B=0$. 
For  $C_1$  positive potential has a minimum at nonzero $B$.
Using the identity
$$({\bf EH})^2={\bf H}^2(2B^2-{\bf H}^2)\cos^2\omega,$$
it is easy to check that there are degenerate absolute minima  corresponding
to field configurations with the parameters 
\begin{eqnarray}
&&{\bf H}^2={\cal B}^2, \ \omega_n=\pi n \ (n=0,1), 
\ \xi_k=(2k+1)\pi/6 \ (k=0,1,\dots,5),
\nonumber\\
&&{\cal B}^2=2\Lambda^4\left(\sqrt{C_2^2+3C_1C_3}-C_2\right)/3C_3>0,
\nonumber
\end{eqnarray} 
These twelve discrete degenerate minima correspond to self-dual and
anti-self-dual field configurations and six 
values of angle $\xi$, and there is a continuous degeneracy
relating to orientations of the chromomagnetic field ${\bf H}$.
With the symplest polynomial form for $U_{\rm eff}$
as above we have $\xi_0=\pi/6$. This value depends on the form 
of effective potential, however the period $\pi/3$  related to the 
Weyl symmetry is universal

\section{The Wilson Loop for $SU(3)$}

For SU(3) the eigenvalues of the colour matrix
\begin{eqnarray}
\hat n_j =t^3\cos\xi_j+t^8\sin\xi_j
\nonumber
\end{eqnarray}
take values from the set
\begin{eqnarray}
\frac{1}{\sqrt{3}},\ -\frac{1}{\sqrt{3}}, \
\frac{1}{2\sqrt{3}},\ -\frac{1}{2\sqrt{3}}
\nonumber
\end{eqnarray}
for all different values of the vacuum angle
\begin{eqnarray}
\xi_j\in((2l+1)\pi/3)_{l=0,\dots,5}.
\nonumber
\end{eqnarray}
This leads to the following result for the trace averaged over
the vacuum angle
\begin{eqnarray}
W(L)
=
\lim_{V,N\to\infty}\left[\int_V\frac{d^4z_j}{V}
\int d\sigma_j\frac{1}{6}
\left(e^{(i/\sqrt{3}) B^j_{\mu\nu}J_{\mu\nu}(z_j)}
+
e^{-(i/\sqrt{3}) B^j_{\mu\nu}J_{\mu\nu}(z_j)}
\right.\right.
\nonumber\\
\left.\left.
+
2e^{(i/2\sqrt{3}) B^j_{\mu\nu}J_{\mu\nu}(z_j)}
+
2e^{-(i/2\sqrt{3}) B^j_{\mu\nu}J_{\mu\nu}(z_j)}
\right)\right]^N.
\nonumber
\end{eqnarray}
Then the Wilson loop takes the form
\begin{eqnarray}
W(L)&=&\lim_{N\to\infty}\left[1-\frac{1}{N}U(L)
\right]^N=e^{-U(L)}
\nonumber\\
U(L)&=&\frac{\pi^2R^2L^2}{3v}
\left(3-
\frac{\sqrt{3}}{2\pi BR^2}\int_0^{2\pi BR^2/\sqrt{3}}\frac{dx}{x}\sin x
-
\frac{2\sqrt{3}}{\pi BR^2}\int_0^{\pi BR^2/\sqrt{3}}\frac{dx}{x}\sin x
\right)
\nonumber\\
&+&\frac{\pi^2}{v}\left(\frac{4}{3}R^3L+\frac{1}{2}R^4\right)
\nonumber\\
&-&\frac{\pi^2(1-\cos2\pi BR^2/\sqrt{3}}{v(2\pi B/\sqrt{3})^2}-
\frac{2\pi^2(1-\cos\pi BR^2/\sqrt{3}}{v(\pi B/\sqrt{3})^2}
\nonumber\\
&+&\frac{4\pi^2L}{v}
\left(\frac{1}{(2\pi B/\sqrt{3})^{3/2}}
\int_0^{\sqrt{2\pi BR^2/\sqrt{3}}}dx\sin x^2 \right. 
\nonumber \\
&+& \left. \frac{2}{(\pi B/\sqrt{3})^{3/2}}
\int_0^{\sqrt{\pi BR^2/\sqrt{3}}}dx\sin x^2
\right)
\nonumber\\
&-&
\frac{\pi^2L^4}{v}
\int_0^{R^2/L^2}ds
\int_{(1-\sqrt{s})^2}^{(1+\sqrt{s})^2}dt
\left(
\sqrt{3}
\frac{\sin\left[
BL^2\left(2\varphi-\sin\varphi+s(2\psi-\sin\psi)
\right)/\sqrt{3}
\right]}{BL^2\left(2\varphi-\sin\varphi+s(2\psi-\sin\psi)
\right)}
\right.
\nonumber\\
&+&\left.
4\sqrt{3}\frac{\sin\left[
BL^2\left(2\varphi-\sin\varphi+s(2\psi-\sin\psi)
\right)/2\sqrt{3}
\right]}{BL^2\left(2\varphi-\sin\varphi+s(2\psi-\sin\psi)
\right)}\right).
\nonumber
\end{eqnarray}
This leads to the string constant as given through the function
as in Eq.(\ref{su(3)f}). 

\section{Scalar Field Eigenvalue Problem}

We introduce the $O(4)$ generators
\begin{eqnarray}
L_i & = & i \epsilon_{ijk} x_j \partial_k,
\nonumber \\
M_i & = & i (x_4 \partial_i - x_i \partial_4),
\nonumber
\end{eqnarray}
respectively for spatial rotations and Euclidean ``boosts''.
These satisfy the usual commutation relations.

Now in the scalar field eigenvalue problem we encounter
the structure
\begin{eqnarray}
(\partial_{\mu} - i B_{\mu})^2 = \partial^2 - 2i B_\mu \partial_\mu
- B_{\mu} B_{\mu}.
\nonumber 
\end{eqnarray}
Given that 
\begin{eqnarray}
B_{\mu} \partial_{\mu} & = & -\frac{1}{2} B_{\mu \nu} x_{\nu} \partial_{\mu}
\nonumber \\
{} & = & \frac{1}{2i} (E M_3 - B L_3)
\nonumber 
\end{eqnarray}
we see that it is better to go over to the $O(3)\times O(3)$ generators,
\begin{eqnarray}
{\bf K}_1 & = & \frac{1}{2} ({\bf L} + {\bf M}) \nonumber \\
{\bf K}_2 & = & \frac{1}{2} ({\bf L} - {\bf M}).
\nonumber
\end{eqnarray}
So with the field self-dual/antiself-dual, $E = \pm B$
we obtain,
\begin{eqnarray}
B_{\mu} \partial_{\mu}  = \left\{ 
        \begin{array}{c} i B K_{2z} , \ E = B \nonumber \\
                         i B K_{1z} , \ E = - B \end{array} \right.  .
\nonumber 
\end{eqnarray}
Then the  four dimensional Laplace operator can be written
\begin{eqnarray}
\partial^2 = {1\over {r^3}} \partial_r (r^3 \partial_r) -
             {4 \over {r^2}} {\bf K}_1^2.
\nonumber
\end{eqnarray}
The usual considerations show then that the eigenvalues of the
complete set of mutually commuting operators ${\bf K}_1^2 = {\bf K}_2^2$,
$K_{1z}$ and $K_{2z}$ are
\begin{eqnarray}
\frac{k}{2}(\frac{k}{2} + 1) \ (k = 0,1,\dots), \ \
m_1,m_2 = \frac{k}{2}, \frac{k}{2} - 1, \dots, - \frac{k}{2}.
\end{eqnarray}
Eigenfunctions corresponding to these values can be found such that
\begin{eqnarray}
K_{1z} C_{k m_1 m_2} & = & m_1 C_{k m_1 m_2} ,
\nonumber\\
K_{2z} C_{k m_1 m_2} & = & m_2 C_{k m_1 m_2},
\nonumber  \\
{\bf K}^2_1 C_{k m_1 m_2} & = & \frac{k}{2} (\frac{k}{2} + 1)
C_{k m_1 m_2}.
\label{evals}
\end{eqnarray}
In the hyperspherical coordinates 
\begin{eqnarray}
x_1 & = & r \sin \eta \cos \phi \nonumber \\
x_2 & = & r \sin \eta \sin \phi \nonumber \\
x_3 & = & r \cos \eta \cos \chi \nonumber \\
x_4 & = & r \cos \eta \sin \chi \nonumber
\end{eqnarray}
$4{\bf K}_1^2$ takes the form,
\begin{eqnarray}
4{\bf K}_1^2 = -{{\partial^2}\over {\partial \eta^2}} 
-{\rm cosec}^2\eta {{\partial^2}\over {\partial \phi^2}} 
-{\rm sec}^2\eta {{\partial^2}\over {\partial \chi^2}}
+ (\tan \eta - \cot \eta) {{\partial}\over {\partial \eta}}.
\nonumber
\end{eqnarray}
Thus
\begin{eqnarray}
4{\bf K}_1^2 C_{k m_1 m_2}(\eta,\phi,\chi) & = & 
\exp i[(m_1-m_2)\chi + (m_1 +m_2)\phi]
\nonumber \\
{}&{}& \times \left[ 
-{{\partial^2}\over {\partial \eta^2}}
+(m_1 + m_2)^2 {\rm cosec}^2\eta 
+(m_1 - m_2)^2 {\rm sec}^2\eta 
\right. \nonumber \\
{}&{}& \left. + (\tan \eta - \cot \eta) {{\partial}\over {\partial \eta}}
\right] \Theta_k^{m_1-m_2,m_1+m_2}(\eta).
\nonumber 
\end{eqnarray}  
Up to normalisation, the angular eigenfunctions $\Theta$
can be better written in terms of the hypergeometric function,
\begin{eqnarray}
\Theta_k^{m_1-m_2,m_1+m_2}(\eta)& \propto & 
\cos^{m_1-m_2}(\eta) \sin^{k-m_1 + m_2}(\eta) 
\nonumber \\
{}&{}& \times
{}_2{\rm F}_1(-\frac{k}{2}+m_1,-\frac{k}{2}-m_2;m_1-m_2+1;-\cot^2\eta).
\nonumber 
\end{eqnarray}
For compactness of notation we denote
\begin{eqnarray}
h(\eta) & = & \cos^{m_1-m_2}(\eta) \sin^{k-m_1 + m_2}(\eta) ,
\nonumber \\
u(\eta) & = & 
{}_2{\rm F}_1(-\frac{k}{2}+m_1,-\frac{k}{2}-m_2;m_1-m_2+1;-\cot^2\eta). 
\nonumber 
\end{eqnarray}
Then, after some tedious calculation, one gets
\begin{eqnarray}
4{\bf K}_1^2 C_{k m_1 m_2}(\eta,\phi,\chi) & = &
\exp i[(m_1-m_2)\chi + (m_1 +m_2)\phi] 
 h(\eta) 
\left[ - \frac{d^2u}{d\eta}  \right. 
\nonumber \\
{}&{}& 
+ \cot \eta \frac{du}{d\eta} \left( {{2(m_1 - m_2)+1}\over{\cot^2\eta}}
                     - (2(k-m_1+m_2)+1) \right)
\nonumber \\
{}&{}& \left.
+ u \left( (2m_1 - k)(2m_2+k) \cot^2\eta + 2k(m_1-m_2+1) +4 m_1 m_2 \right).
\right] 
\nonumber 
\end{eqnarray}
Rewriting in terms of the variable $z=-\cot^2\eta$
one eventually brings this to the form  
\begin{eqnarray}
4{\bf K}_1^2 C_{k m_1 m_2}(\eta,\phi,\chi) & = &
\exp i[(m_1-m_2)\chi + (m_1 +m_2)\phi] 
\times 4 h(\eta) (1-z) 
\nonumber \\
{}&{}& \times 
\left[ z(1-z) \frac{d^2u}{dz^2} 
+ \frac{du}{dz} \left( (m_1-m_2 + 1) - (1-k+m_1-m_2) z \right) 
\right. \nonumber \\
{}&{}& \left. 
- (\frac{k}{2} - m_1) (\frac{k}{2} + m_2) u 
+\frac{k}{2}(\frac{k}{2}+1) {{u}\over{1-z}} \right].
\label{2ndlast}
\end{eqnarray}
But $u$ satisfies the hypergeometric equation. Thus the
first three set of terms in the square bracket of Eq.(\ref{2ndlast}) vanish,
leaving
\begin{eqnarray}
4{\bf K}_1^2 C_{k m_1 m_2}(\eta,\phi,\chi)  = 
4\frac{k}{2}(\frac{k}{2}+1) e^{ i[(m_1-m_2)\chi + (m_1 +m_2)\phi]}
h(\eta) u(\eta), 
\nonumber 
\end{eqnarray}
namely, Eq.(\ref{evals}), which completes the proof.

Putting all this together we arrive at the following
representation for the square of the covariant derivative
\begin{eqnarray}
(\partial_{\mu} - i B_{\mu})^2 &=& {1\over {r^3}} \partial_r (r^3 \partial_r) -
             {4 \over {r^2}} {\bf K}_1^2
            + 2B K_{2z} - \frac{1}{4} B^2 r^2
, \ E = B \nonumber \\
(\partial_{\mu} -i B_{\mu})^2  &=& {1\over {r^3}} \partial_r (r^3 \partial_r) -
             {4 \over {r^2}} {\bf K}_1^2
            + 2B K_{1z} - \frac{1}{4} B^2 r^2
, \ E = -B.
\nonumber
\end{eqnarray}
Using the eigenfunctions $C_{km_1m_2}$ we can reduce the
original eigenvalue problem to that corresponding to a radial operator
\begin{eqnarray}
- [ {1\over {r^3}} \partial_r (r^3 \partial_r) -
{ {k(k+2)} \over {r^2} } + 2 m_{2,1} B -
{ {B^2 r^2} \over 4} - {\cal M}^2 ] f(r) = \lambda f(r),
\nonumber
\end{eqnarray}
where $m_{2,1}$ is defined in the main text.

So the complete eigenfunctions will be a product of radial and
angular parts,
\begin{eqnarray}
\psi(x) = f(r) C_{k m_1 m_2}(\eta,\phi,\chi).
\nonumber 
\end{eqnarray}
The radial function can be solved, as described in the main
body of the paper, leading to the solution given in
Eq.(\ref{radialsol}).

A comment at this point on the half-integer values of the azimuthal
quantum numbers. The angular eigenfunctions depend only on the
sum and differences of $m$s, which will be whole integers. Also
the combination
\begin{eqnarray}
n_{2,1} \equiv \frac{k}{2} - m_{2,1} + 1
& = & \frac{k}{2} + \frac{k}{2} +1, \frac{k}{2} +
\frac{k}{2} ,
\dots, \frac{k}{2} - \frac{k}{2} + 1\nonumber \\
        {}      & = & k+1, k, \dots, 1.
\nonumber
\end{eqnarray}
This combination, and thus the eigenvalue spectrum, will always
involve integral values.

\section{Derivation of Scalar Field Propagator}

We represent the delta function in the hyperspherical coordinates,
$$
\delta^{(4)}(x-x') =
{ {\delta(r-r') \delta(\eta - \eta') \delta(\phi-\phi') \delta(\chi-\chi')}
\over {r^3 \sin \eta \cos \eta} },  
$$
where the primed variables correspond to the hyperspherical coordinates
of $x'$.  We use the ansatz,
$$
G(x,x'|\mu) = \sum_{k,m_1,m_2} V_{k m_1 m_2}(\eta',\phi',\chi')
C_{k m_1 m_2}(\eta,\phi,\chi) X_{k m_1 m_2} (r,r'|\mu). \nonumber 
$$
Inserting this into the original Green's function equation,
exploiting the delta-function representation
and the fact that the functions $C$ are eigenfunctions of
the covariant derivative squared operator, yields
\begin{eqnarray}
{} & {} & \sum_{k,m_1,m_2} V_{k m_1 m_2}(\eta',\phi',\chi')
C_{k m_1 m_2}(\eta,\phi,\chi)
  \nonumber \\
{} & {} & \times
\left[ {1\over {r^3}} \partial_r (r^3 \partial_r) -
                  { {k(k+2)} \over {r^2} } + 2 m_{2,1} B -
 {{B^2 r^2} \over 4} - {\cal M}^2 \right]
                                X_{k m_1 m_2}(r,r'|\mu),
   \nonumber \\
{}  & {} & \hspace{5cm}  =  -
{ {\delta(r-r') \delta(\eta - \eta') \delta(\phi-\phi') \delta(\chi-\chi')}
\over {r^3 \sin \eta \cos \eta} }.
\nonumber 
\end{eqnarray}
We can now separate the angular from the radial dependence
in two equations,
\begin{eqnarray}
{} & {} &
\sum_{k,m_1,m_2} V_{k m_1 m_2} (\eta',\phi',\chi')
C_{k m_1 m_2}(\eta,\phi,\chi)  =
{ {\delta(\eta - \eta') \delta(\phi-\phi') \delta(\chi-\chi')}
\over { \sin \eta \cos \eta} }, \label{angeqn} \\
{} & {} &
\left[ {1\over {r^3}} \partial_r (r^3 \partial_r) -
                  { {k(k+2)} \over {r^2} } + 2 m_{2,1} B -
 {{B^2 r^2} \over 4} - {\cal M}^2 \right]
                                X_{k m_1 m_2}(r,r'|\mu)
  =  - { {\delta(r-r')} \over {r^3} }.
\nonumber \\
\label{radeqn}
\end{eqnarray}
Equation (\ref{angeqn}) we recognise as the completeness
relation for the angular eigenfunctions, thus $V = C$.
Using Eq.(\ref{radeqn}), we read off that $X$ does not
depend on both quantum numbers $m_2,m_1$ but on one
of them, depending on whether the field was self-dual
or anti-self-dual. In fact we shall indicate this
dependence on $m_{2,1}$ via the quantum number $n_{2,1}$
so that $X= X_{k n_{2,1}}(r,r'|\mu)$.

We can solve the radial problem by solving for the radial
Green's function in the infinite volume and then adding
a solution to the homogeneous equation with an arbitrary
coefficient. The coefficient is fixed by imposing the finite
boundary condition at $r=R$.

Using our solution to the eigenvalue equation we can easily
extract homogeneous solutions, namely to the equation:
$$
\left[ {1\over {r^3}} \partial_r (r^3 \partial_r) -
                  { {k(k+2)} \over {r^2} } + 2 m_{2,1} B -
 {{B^2 r^2} \over 4} - {\cal M}^2 \right] R(r) = 0.
$$
In the notation of \cite{AS65},
the two solutions (respectively regular at infinity and the origin) are
\begin{eqnarray}
R_1(r|k,n_{2,1},\mu) & = &
r^k e^{-Br^2/4} M(n_{2,1} + \mu,
k+2; Br^2/2) \nonumber \\
R_2(r|k,n_{2,1},\mu) & = &
r^k e^{-Br^2/4} U(n_{2,1} + \mu,
k+2; Br^2/2)
\end{eqnarray}
with $n_{2,1}$ defined in Eq.(\ref{index}).

We next solve for the Green's function in the $R=\infty$ case by
recasting the problem in the
form of a Sturm-Liouville equation,
$$
\left[ \partial_r p(r) \partial_r + q(r) \right] X(r,r') =
- \delta(r-r'),
$$
so that
\begin{eqnarray}
p(r) & = & r^3 \nonumber \\
q(r) & = & - \left[
   { {B^2 r^2}\over 4} + {{k(k+2)}\over {r^2}} - 2B m_{2,1} + {\cal M}^2
\right] r^3
\nonumber
\end{eqnarray}
The Sturm-Liouville equation is known to have solution,
$$
X(r,r') = - {1 \over { p(r') w(r') } }
R(r,r')
$$
with $w(r)$ the Wronskian of the homogeneous solutions,
$$
w = R_1 R_2^{\prime} - R_2 R_1^{\prime}
$$
and
\begin{eqnarray}
R(r,r')  = \left\{ 
\begin{array}{c}  R_1(r) R_2(r'), \ r < r' \\
                  R_1(r') R_2(r), \ r > r'
\end{array} . \right. .
\nonumber 
\end{eqnarray}

The Wronskian for the two solutions is evaluated to be
$$
w(r) = - { {\Gamma(k+2) \sqrt{2B} }  \over
 { \Gamma(n_{2,1} + \mu) (Br^2/2)^{3/2} }  }
$$
so that
$$
p(r) w(r) = { { -4 \Gamma(k+2) } \over
   { B \Gamma(n_{2,1} + \mu) } } .
$$

We now construct the full Green's function and impose the
boundary condition at finite R:
\begin{eqnarray}
G(x,x'|\mu) & = & \sum_{k,m_1,m_2} C_{k m_1 m_2}(\eta',\phi',\chi')
C_{k m_1 m_2}(\eta,\phi,\chi)
\left[ X_{k n_{2,1}}(r,r'|\mu) \right.   \nonumber \\
  & {} & \left. +  A_{k n_{2,1}}(R|B,\mu)
R_1(r|k, n_{2,1},\mu) R_1(r'|k, n_{2,1},\mu)
\right].
\nonumber
\end{eqnarray}
Imposing $G(R,r'|\mu)=0$ with $r'<r=R$ we have
\begin{eqnarray}
A_{k n_{2,1}}(R|\mu) & = & - {{X_{k n_{2,1}}(R,r'|\mu)} \over
                { R_1(R|k,n_{2,1},\mu) R_1(r'|k,n_{2,1},\mu) }  }
\nonumber \\
   {} & = & + { {R_{k n_{2,1}}(R,r')} \over
 {w(R)p(R) R_1(R|k,n_{2,1},\mu) R_1(r'|k,n_{2,1},\mu) }  } \nonumber \\
   {} & = & + { {R_1(r'|k,n_{2,1},\mu) R_2(R|k,n_{2,1},\mu)} \over
 {w(R)p(R) R_1(R|k,n_{2,1},\mu) R_1(r'|k,n_{2,1},\mu) }  } \nonumber \\
   {} & = & { {R_2(R|k,n_{2,1},\mu)} \over {w(R)p(R) R_1(R|k,n_{2,1},\mu)}}.
\nonumber
\end{eqnarray}
Thus finally,
\begin{equation}
A_{k n_{2,1}}(R|\mu)=-B{{\Gamma(n_{2,1} + \mu)}
\over {4 \Gamma(k+2)} }
{ {U(n_{2,1} + \mu, k+2;BR^2/2)}
\over
  {M(n_{2,1} + \mu, k+2;BR^2/2)} } .
\nonumber \\
\end{equation}
It is straightforward to see that imposing $G(r,R|\mu)=0, \ r < R$
will give the same result for $A$.

\end{document}